\documentclass[a4paper,fleqn,usenatbib]{mnras}

\usepackage{newtxtext,newtxmath}

\usepackage[T1]{fontenc}
\usepackage{ae,aecompl}

\usepackage{graphicx}	
\usepackage{amsmath}	
\usepackage{amssymb}	
\usepackage{CJKutf8}	
\usepackage{multirow}   

\newcommand{\herschel}{{\it Herschel\/}}       
\newcommand{\chandra}{{\it Chandra\/}}

\newcommand{\xmm}{\hbox{\it XMM-Newton\/}}

\newcommand{\athena}{{\it Athena\/}}

\newcommand{\spitzer}{{\it Spitzer\/}}

\newcommand{\akari}{{\it AKARI\/}}  
\newcommand{\erosita}{{\it eROSITA\/}}  

\newcommand{\xray}{\hbox{X-ray}}  

\newcommand{\mbh}{M_{\rm BH}}
\newcommand{\mstar}{M_{\star}}
\newcommand{\ox}{\alpha_{\rm ox}}
\newcommand{\ang}{\textup{\AA}}
\newcommand{\luv}{L_{2500 \ang}}
\newcommand{\luvx}{L_{2500 \ang, \rm X}}
\newcommand{\luvnox}{L_{2500 \ang, \rm noX}}
\newcommand{\fracA}{{\rm frac_{AGN}}}
\newcommand{\fracAx}{{\rm frac_{AGN, Xup}}}
\newcommand{\fracAnox}{{\rm frac_{AGN, noX}}}
\newcommand{\lxr}{L_{\rm 2 keV}}
\newcommand{\xcig}{\hbox{X-CIGALE}}
\newcommand{\redchi}{\chi^2_{\rm red}}

\title[X-CIGALE]{X-CIGALE: Fitting AGN/galaxy SEDs from X-ray to infrared}

\author[G. Yang et al.]{
G. Yang\begin{CJK*}{UTF8}{gbsn} (杨光),\end{CJK*}$^{1,2,3}$\thanks{E-mail: gyang206265@gmail.com (GY)}
M. Boquien,$^4$
V. Buat,$^3$
D. Burgarella,$^3$
L. Ciesla,$^3$
F. Duras,$^3$
\newauthor
M. Stalevski,$^{5,6}$
W. N. Brandt,$^{7,8,9}$
and
C. Papovich$^{1,2}$
\\
$^{1}$Department of Physics and Astronomy, Texas A\&M University, College Station, TX 77843-4242, USA \\
$^{2}$George P. and Cynthia Woods Mitchell Institute for Fundamental Physics and Astronomy, Texas A\&M University, College Station, TX 77843-4242, USA\\
$^{3}$Aix Marseille Univ, CNRS, CNES, LAM, Marseille, France\\
$^{4}$Centro de Astronom\'ia (CITEVA), Universidad de Antofagasta, Avenida Angamos 601, Antofagasta, Chile \\
$^{5}$Astronomical Observatory, Volgina 7, 11060 Belgrade, Serbia \\
$^{6}$Sterrenkundig Observatorium, Universiteit Gent, Krijgslaan 281-S9, Gent, 9000, Belgium \\
$^{7}$Department of Astronomy and Astrophysics, 525 Davey Lab, The Pennsylvania State University, University Park, PA 16802, USA\\
$^{8}$Institute for Gravitation and the Cosmos, The Pennsylvania State University, University Park, PA 16802, USA\\
$^{9}$Department of Physics, 104 Davey Laboratory, The Pennsylvania State University, University Park, PA 16802, USA
}

\date{Accepted XXX. Received YYY; in original form ZZZ}

\pubyear{2018}

\begin{document}
\label{firstpage}
\pagerange{\pageref{firstpage}--\pageref{lastpage}}
\maketitle

\begin{abstract}
CIGALE is a powerful multiwavelength spectral energy distribution 
(SED) fitting code for extragalactic studies. 
However, the current version of CIGALE is not able to fit \xray\ 
data, which often provide unique insights into AGN intrinsic power. 
We develop a new \xray\ module for CIGALE, allowing it to fit SEDs 
from the \xray\ to infrared (IR).
We also improve the AGN fitting of CIGALE from UV-to-IR wavelengths. 
We implement a modern clumpy two-phase torus model, SKIRTOR.
To account for moderately extincted type~1 AGNs, we implement 
polar-dust extinction.
We publicly release the source code (named ``\xcig'').
We test \xcig\ with \xray\ detected AGNs in SDSS, COSMOS, and 
\hbox{AKARI-NEP}.
The fitting quality (as indicated by reduced $\chi^2$) is good 
in general, indicating that \xcig\ is capable of modelling 
the observed SED from \xray\ to IR.
{We discuss constrainability and degeneracy of model 
parameters in the fitting of \hbox{AKARI-NEP}, for which excellent 
mid-IR photometric coverage is available.}
We also test fitting a sample of \hbox{AKARI-NEP} galaxies for 
which only \xray\ upper limits are available from \chandra\ 
observations, 
and find that the upper limit can effectively constrain the AGN 
SED contribution for some systems.
{Finally, using \xcig, we assess the ability of \athena\
to constrain the AGN activity in future extragalactic studies.}
\end{abstract}




\section{Introduction}\label{sec:intro}

Supermassive black holes (BHs) commonly exist in the centers of 
massive galaxies (e.g. \hbox{\citealt{kormendy95}}; 
\hbox{\citealt{kormendy13}}).
BHs grow their mass ($\mbh$) by accreting local material.
During this process, a significant amount of the gravitational 
energy of the accreted material is converted to radiation, and 
the system shines as an active galactic nucleus (AGN).
The typical spectral energy distribution (SED) of AGNs covers 
a broad wavelength range, from \xray\ to infrared (IR).

AGN emission at different wavelengths is generated by different
physical processes \citep[e.g.][]{netzer13}. 
The accretion disk mostly produces photons at ultraviolet (UV)
and optical wavelengths.
Some of these photons are scattered to \xray\ energies by the hot 
corona above the disk (i.e. inverse Compton scattering).
Some of the UV/optical photons might also be absorbed by dust.
The dust is thus heated and reemits the energy as infrared radiation.
Considering the tight link between AGN multiwavelength SEDs and 
these physical processes, it is feasible to infer source properties 
from modelling the observed photometric data. 
On the other hand, the observed SED is often complicated,
involving factors such as host-galaxy contributions and dust extinction.
Misinterpretation of the SED could lead to unrealistic physical 
properties. 
Therefore, it is critical to decipher the observed data appropriately
with a powerful and reliable SED fitting code.

The Code Investigating GALaxy Emission (CIGALE) is a state-of-the-art 
python code for SED fitting of extragalactic sources \citep{boquien19}. 
It employs physical AGN and galaxy models, and allows flexible 
combination between them.
The current version of CIGALE can simultaneously fit the observed 
SED from UV to far-IR (FIR) and extract source physical properties 
such as AGN luminosity and host stellar mass ($\mstar$).
However, the current CIGALE is not able to model \hbox{X-ray}
fluxes, which often provide a unique view of AGNs.

X-ray observations have many advantages in AGN studies 
(see \citealt{brandt15} for a review).
Strong \xray\ emission is nearly a universal property of the 
AGN phenomenon.
\hbox{X-rays} are generated from the immediate vicinity of the 
BH, directly revealing the intrinsic AGN power.
Therefore, \xray\ fluxes are widely used as a tracer of BH
accretion rate \citep[e.g.][]{yang18, yang19}.
Thanks to their great penetrating power, \hbox{X-rays} are only
mildly affected by obscuration in general.
Also, AGNs are much more efficient in generating \hbox{X-rays}
than their host galaxies. 
Therefore, the observed \xray\ fluxes are often dominated by 
AGNs and have negligible galaxy contribution. 
Considering these advantageous properties, \xray\ observations are 
widely used to select AGNs, especially in the distant universe, 
\citep[e.g.][]{luo17, chen18}.
These selections are often more complete and reliable than the 
selections at other wavelengths such as optical and IR.

Besides the lack of \xray\ fitting capability, CIGALE's current 
AGN model \citep{fritz06}, which covers the UV to IR, also 
has some other disadvantages.
The model assumes that the central engine is surrounded by a dusty 
torus (i.e. the AGN unified model; \hbox{\citealt{antonucci93}}; 
\hbox{\citealt{urry95}}; \hbox{\citealt{netzer15}}; 
\hbox{\citealt{zou19}}).
The torus absorbs a fraction of the UV and optical emission from
the central engine and reemits the energy as IR photons.
When viewing from the equatorial direction, the central engine is 
obscured and only reemitted IR radiation can be observed
(type~2 AGN).
When viewing from the polar direction, the central engine is 
directly visible (type~1 AGN).

One disadvantage of the AGN model is that it assumes the dusty torus 
is a smooth structure.
However, such smooth models for the torus are disfavored on 
physical grounds \citep[e.g.][]{tanimoto19}. 
To reach a scale height consistent with observations, the dust 
grains in a smooth torus would have random velocities 
$\sim 100$~km~s$^{-1}$, corresponding to a temperature of 
$\sim 10^6$~K. 
This high temperature far exceeds the dust-sublimation temperature
($\sim 10^3$~K).
Another disadvantage of the AGN model is that the disk emission is 
assumed to be absolutely unextincted for the case of type~1.
{However, recent observations indicate that a non-negligible amount 
of extinction exists for some type~1 AGNs (e.g. 
\hbox{\citealt{bongiorno12}}; \hbox{\citealt{elvis12}};
\hbox{\citealt{lusso12}}), 
which can be attributed to the dust existing along polar directions (e.g. 
\hbox{\citealt{stalevski17, stalevski19}}; \hbox{\citealt{lyu18}}).}
The current CIGALE cannot model the SEDs of these type~1 AGNs.

In this paper, we further develop CIGALE and enable it to fit \xray\ data. 
The new development allows CIGALE to model AGN SED from \xray\ to IR
simultaneously and extract source properties such as AGN intrinsic 
luminosity and host-galaxy stellar mass ($\mstar$).
Besides developing the \xray\ part, we also improve CIGALE's 
capability in fitting the UV-to-IR SED of AGNs.
We implement the latest version of SKIRTOR, a clumpy two-phase torus 
model derived from a modern radiative-transfer method \citep{stalevski12, 
stalevski16}.
In addition, we introduce polar-dust extinction to account for the possible 
extinction in type~1 AGNs.
We name the new version of CIGALE as ``\xcig''. 

The structure of this paper is as follows.
In \S\ref{sec:code}, we outline the scheme of our new code development.
In \S\ref{sec:test}, we test \xcig\ on AGNs with \xray\ detections 
from different surveys.
We test fitting galaxies with only \xray\ upper limits
in \S\ref{sec:uplim}.
We summarize our results and discuss future prospects in 
\S\ref{sec:sum}.

Throughout this paper, we assume a flat $\Lambda$CDM cosmology with 
$H_0=69.3$~km~s$^{-1}$~Mpc$^{-1}$ and $\Omega_M=0.286$
(WMAP 9-year results; \hbox{\citealt{hinshaw13}}).
Quoted uncertainties are at the $1\sigma$\ (68\%) confidence level, unless
otherwise stated.
Quoted optical/infrared magnitudes are AB magnitudes.

\section{The code}
\label{sec:code}
We briefly summarize the mechanisms and features of CIGALE in 
\S\ref{sec:cigale}.
In \S\ref{sec:xray_mod}, \S\ref{sec:skirtor}, and 
\S\ref{sec:polar_dust}, we detail our new development of \xcig, 
i.e. the \xray\ fitting, SKIRTOR, and the polar-dust extinction.
The new inputs/outputs introduced in \xcig\ are listed in 
Appendix~\ref{sec:in_and_out}.

\subsection{A brief introduction of CIGALE}\label{sec:cigale}
CIGALE is an efficient SED-fitting code which has been developed 
for more than a decade 
(\hbox{\citealt{burgarella05}}; \hbox{\citealt{noll09}}; 
\hbox{\citealt{serra11}}; \hbox{\citealt{boquien19}}).
CIGALE is written in Python. 
{\xcig\ is built upon CIGALE, and the fitting algorithm 
of \xcig\ is the same as that of CIGALE.
Here, we only briefly introduce the algorithm, and interested 
readers should refer to \cite{boquien19} for a detailed 
description.}

CIGALE allows the user to input a set of model parameters. 
The code then realizes the model SED for each possible combination 
of the model parameters, and convolves the model SED with the filters 
to derive model fluxes.
By comparing the model fluxes with the observed fluxes, 
{the code computes likelihood as $L = \exp(-\chi^2/2)$ for 
each model.
CIGALE supports two types of analyses, i.e. 
maximum likelihood (minimum $\chi^2$) and {Bayesian-like}. 
In the maximum-likelihood analyses, CIGALE picks out the model 
with the largest $L$ value, and calculates physical properties 
such as $\mstar$ and star formation rate (SFR) from this single 
model.
In the {Bayesian-like} analyses, for each physical property, CIGALE 
calculates the marginalized probability distribution function 
(PDF) based on the $L$ values of all models. 
Finally, from this PDF, CIGALE derives the probability-weighted 
mean and standard deviation, and outputs them as the estimated 
value and uncertainty.
}

Among the above processes, one key step is the realization of model 
SEDs from input parameters. 
This procedure relies on a set of modules, and each module is 
responsible for a function that shapes the SED. 
For example, the ``nebular emission'' module adds the nebular-emission
components to the SED, and the ``dust attenuation'' module extincts 
the SED.
Our new development of \xcig\ follows this module-based structure. 
We enable CIGALE to fit \xray\ data by developing a new \xray\ module 
(\S\ref{sec:xray_mod});
we implement SKIRTOR templates and polar-dust extinction in a new 
SKIRTOR module (\S\ref{sec:skirtor} and \S\ref{sec:polar_dust}).

\subsection{The new X-ray module}\label{sec:xray_mod}
In this section, we develop a new \xray\ module to enable \xcig\ 
to fit \xray\ data.
In \S\ref{sec:set}, we detail the basic settings of this new 
module.
In \S\ref{sec:xray_sed}, we present the adopted \xray\ SED for AGN
and galaxy components. 
In \S\ref{sec:ox}, we present the relation that we used to link
AGN \xray\ with other wavelengths.
We note that our new developments are for the majority AGN 
population in optical/\xray\ surveys, and thus \xcig\ may not be 
applicable to some minor populations such as radio-loud and broad 
absorption line (BAL) objects \citep[e.g.][]{brandt00, miller11}.
We leave the treatment of these particular AGNs to future works.

\subsubsection{Basic settings}\label{sec:set}
As presented in \S\ref{sec:intro}, the \xray\ band has many 
advantages in studying AGNs. 
Therefore, we implement an \xray\ module for \xcig.
The main goal of this module is to connect \xray\ with other 
wavelengths, rather than to obtain detailed \xray\ spectral 
properties (e.g. photon index and hydrogen column density) by 
performing detailed \xray\ spectral analyses. 
This is because the latter has already been well realized by 
many specialized \xray\ codes such as XSPEC \citep{arnaud96} 
and Sherpa \citep{freeman01}, and there is no need for \xcig\ 
to perform similar analyses.
Also, it is technically difficult to fit the \xray\ spectra within 
the framework of \xcig. 
\xcig\ assumes that a sample of sources are observed with a single 
``filter transmission'', as is the case in UV-to-IR data. 
However, at \xray\ wavelengths, the transmission curve varies from 
source-to-source, as it might depend on many factors such as position 
on the detectors and observation date.
For example, the soft-band transmission of \chandra\ has been 
continuously declining since its launch \citep[e.g.][]{odell17}.
In fact, each source is associated with a unique transmission curve
and the curve is taken into account when fitting the \xray\ spectra 
with, e.g. XSPEC and Sherpa.

Therefore, the \xray\ module of \xcig\ is designed to work on the 
high-level \xray\ data products, i.e. intrinsic \xray\ fluxes in a 
given band.
We require the \xray\ fluxes to be corrected for telescope transmission. 
Fortunately, this correction is embedded in routine \xray\ data 
processing and has already been applied in \xray\ photometric catalogs
\citep[e.g.][]{yang16, luo17}. 
Since the transmission has already been considered, \xcig\ only needs 
to adopt a uniform-sensitivity (i.e. boxcar-shaped) ``filter''.
We have already included a few typical boxcar \xray\ filters, e.g. 
\hbox{0.5--2~keV} and \hbox{2--7~keV} for convenience, while
the user can easily generate the filters for any \xray\ band.

In addition, we require the input \xray\ fluxes to be 
``absorption-corrected''.
The absorption might be from the source itself, {our Galaxy}, 
and/or the intergalactic medium (IGM; e.g. \hbox{\citealt{starling13}}). 
However, we do not differentiate these types of absorption, as 
it is often infeasible to separate them in \xray\ data analyses.
The absorption correction can be obtained from routine \xray\ data
processing, e.g., spectral analyses via XSPEC/Sherpa or band-ratio 
analyses \citep[e.g.][]{xue16, luo17}.
The user may also choose to use hard \xray\ bands where 
absorption corrections are generally small 
\citep[e.g.][]{yang18, yang18b}.

In \xray\ catalogs \citep[e.g.][]{xue16, luo17}, \xray\ fluxes 
($f_{\rm X}$) are conventionally given in the cgs units of 
erg~s$^{-1}$~cm$^{-2}$, but \xcig\ requires the input fluxes 
($f_{\rm cigale}$) to be given in the units of mJy. 
Therefore, the user needs to convert the flux units with 
\begin{equation}
f_{\rm cigale} = \frac{f_{\rm x} \times 4.136 \times 10^8}{E_{\rm up} - E_{\rm lo}}
\end{equation}
where $E_{\rm lo}$ and $E_{\rm up}$ refer to the lower and upper
limits of the energy band in units of keV. 

The \xray\ module covers rest-frame \hbox{$10^{-3}$--5 nm}, 
corresponding to \hbox{$\approx$~0.25--1200 keV}.
Such an energy range is sufficient for practical purposes:
current \xray\ instruments cannot observe energies significantly 
below rest-frame $0.5(1+z)$~keV in general;
the AGN flux is typically non-detectable above $\approx 1000$~keV
due to the existence of the cut-off energy in AGN \xray\ spectra  
(see \S\ref{sec:xray_sed}).

\subsubsection{X-ray SED}
\label{sec:xray_sed}
\begin{figure}
\includegraphics[width=\columnwidth]{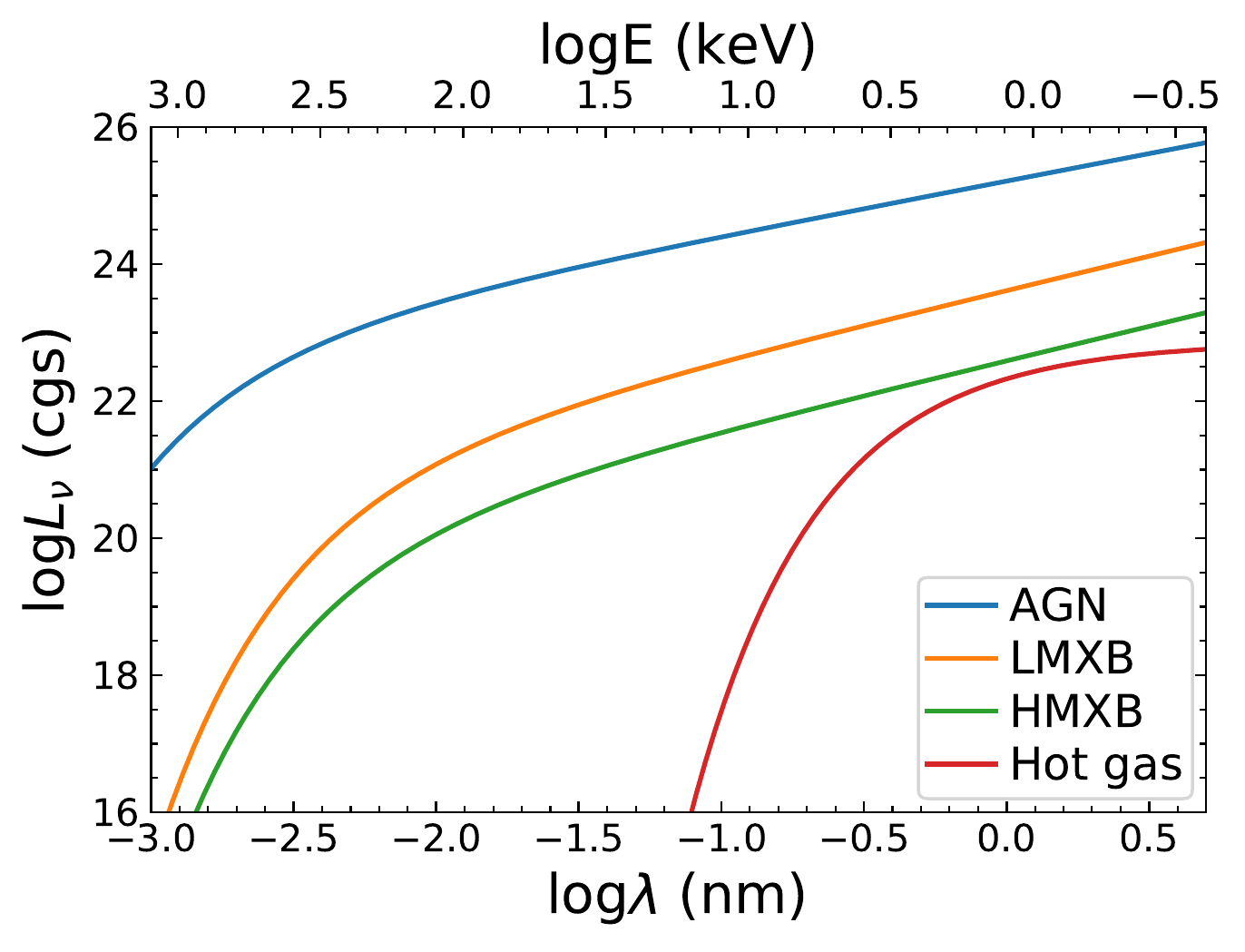}
\caption{An example \xray\ SED model for a typical source 
with AGN $L_{\rm 2\text{--}10\ keV} = 10^{43}$~erg~s$^{-1}$, 
$\mstar = 10^{11}$~$M_\odot$, ${\rm SFR} = 10\ M_\odot$~yr$^{-1}$,
$T = 1$~Gyr, and $Z = 0.02$.  
Different colors indicate different components.
For this source, the \xray\ luminosity is dominantly contributed
by the AGN.
}
\label{fig:xray_sed}
\end{figure}

To first-order approximation, the intrinsic AGN \xray\ spectrum 
is typically a power law with a high-energy exponential cutoff, i.e.
\begin{equation}
\label{eq:agn_sed}
    f_\nu \propto E^{-\Gamma + 1} \exp(-E/E_{\rm cut})
\end{equation}
where $\Gamma$ is the so-called ``photon index'', widely adopted 
in \xray\ astronomy, and $E_{\rm cut}$ is the cutoff energy.
We adopt this spectral shape in \xcig.
Detailed \xray\ spectral fitting in the literature finds
$\Gamma \approx 1.8$ \citep[e.g.][]{yang16, liu17}.
We allow the user to set $\Gamma$ in \xcig.
We set $E_{\rm cut}=300$~keV, the typical value from the 
observations of Seyferts \citep[e.g.][]{dadina08, ricci17}.
Note that since $E_{\rm cut}$ is above the highest observable
energy of most \xray\ observatories (e.g. \chandra\ and \xmm),
the exact choice of $E_{\rm cut}$ has practically negligible 
effects on the fitting with \xcig\ for most cases.
The adopted AGN \xray\ SED is displayed in Fig.~\ref{fig:xray_sed}.

Besides AGNs, galaxies can also emit \hbox{X-rays}, although the 
emission from galaxies is often much weaker than that from AGNs
for \xray\ detected sources.
There are three main origins of galaxy \xray\ emission:
low-mass \xray\ binaries (LMXB), high-mass \xray\ binaries (HMXB), 
and hot gas.
The strengths of these components can be modeled as a function 
of galaxy properties such as $\mstar$ and SFR.
We adopt the recipe from \cite{mezcua18}, where a 
\cite{chabrier03} initial mass function (IMF) is assumed.
In this scheme, the LMXB and HMXB luminosities (in units of erg~s$^{-1}$) 
are described as
\begin{equation}
\begin{split}
\log(L_{\rm 2-10\ keV}^{\rm LMXB}/\mstar) = 40.3 - 1.5\log t - 0.42(\log t)^2 + \\ 
                              0.43(\log t)^3 + 0.14(\log t)^4 \\
\log(L_{\rm 2-10\ keV}^{\rm HMXB}/{\rm SFR}) = 40.3 - 62Z + 569Z^2 - 1834Z^3 + 1968Z^4
\end{split}
\end{equation}
where $\mstar$ and SFR are in solar units; $t$ denotes stellar age 
in units of Gyr; $Z$ denotes metallicity (mass fraction). 
The hot-gas luminosity (in units of erg~s$^{-1}$) is described as
\begin{equation}
\log(L_{\rm 0.5-2 keV}^{\rm hot gas}/{\rm SFR}) = 38.9
\end{equation}
Similarly as for AGN, we also employ the SED shape in Eq.~\ref{eq:agn_sed}
for all three components, {with $E_{\rm cut}$ fixed at 
100~keV (LMXB and HMXB; e.g. \hbox{\citealt{zhang97}}; \hbox{\citealt{motta09}}) 
and 1~keV (hot gas; e.g. \hbox{\citealt{mathews03}}).
We allow the user to set $\Gamma$ values for the LMXB and HMXB components.
In our test fitting in \S\ref{sec:test}, we set $\Gamma$ to 1.56 and 2.0 for 
LMXB and HMXB, respectively \citep[e.g.][]{fabbiano06, sazonov17}.
Adjusting these $\Gamma$ does not affect the fitting results significantly, 
as the observed \xray\ fluxes are often dominantly contributed 
by AGNs rather than galaxies.
The \xray\ continuum from hot gas can be modelled as free-free and free-bound 
emission from optically thin plasma ($\Gamma=1$; e.g. \hbox{\citealt{mewe86}}). 
Therefore, we fix $\Gamma=1$ for the hot-gas component in \xcig.
}
Fig.~\ref{fig:xray_sed} shows the adopted \xray\ SEDs of the 
three components.
We add all three components for the total \xray\ SED 
from galaxies. 

\subsubsection{The $\ox$-$\luv$ relation}
\label{sec:ox}
As in \S\ref{sec:set}, the main goal of \xcig\ is to fit
\xray\ and other wavelengths simultaneously.
Some known connections between \xray\ and other wavelengths must 
be applied; otherwise, the fitting would be practically useless.
We adopt the well-studied ``$\ox$-$\luv$'' relation 
\citep[e.g.][]{steffen06, just07, lusso17}, 
where $\luv$ is AGN intrinsic (de-reddened) 
luminosity per frequency at 2500~\AA\ and $\ox$ is the SED
slope between UV (2500~\AA) and \xray\ (2~keV), i.e.
\begin{equation}
\ox = -0.3838 \log( \luv/\lxr ).
\end{equation}
The observed $\ox$-$\luv$ relation \citep{just07} is written as 
\begin{equation}
\ox = -0.137 \log( \luv ) + 2.638
\end{equation}
where $\luv$ is in units of erg~s$^{-1}$~Hz$^{-1}$.
The 1$\sigma$ intrinsic dispersion of this $\ox$-$\luv$ relation 
is $\Delta \ox \approx 0.1$ (see Table~8 of \hbox{\citealt{just07}}).
Here, $\Delta \ox$ is the $\ox$ deviation from that expected 
from the $\ox$-$\luv$ relation, i.e.
\begin{equation}
\Delta \ox = \ox - \ox(\luv).
\end{equation}
Observations have found that the $\ox$-$\luv$ relation does 
not have significant redshift evolution, indicating that the 
relation originates from fundamental accretion physics
\citep{steffen06, just07, lusso17}.
We allow the user to set the maximum $|\Delta \ox|$ allowed 
($|\Delta \ox|_{\rm max}$).
Internally, \xcig\ calculates all models with $\ox$ from 
$-1.9$ to $-1.1$ with a step of 0.1.\footnote{This $\ox$ 
range corresponds to \hbox{2--10 keV} \xray\ bolometric 
corrections ranging from $\approx 10$ to $\approx 500$.
}
\xcig\ then calculates $|\Delta \ox|$ and {discards}
the models with $|\Delta \ox| > |\Delta \ox|_{\rm max}$.
In our test fitting (\S\ref{sec:test} and \S\ref{sec:uplim}),
we adopt $|\Delta \ox|_{\rm max} = 0.2$, corresponding to 
the $\approx 2\sigma$ scatter of the $\ox$-$\luv$ relation
\citep{just07}.

Note that the $\ox$-$\luv$ relation above is derived from 
observations, assuming that the unobscured AGN emission 
is isotropic at both UV/optical and \xray\ wavelengths.  
However, the UV/optical emission is unlikely isotropic, 
because it is from the accretion disk and the effects of 
projected area and limb darkening affect the angular 
distribution of the radiative energy.
After considering these effects, the disk luminosity can
be approximated as 
$L(\theta) \propto \cos \theta(1+2 \cos \theta)$, 
where $\theta$ is the angle from the AGN axis 
\citep[e.g.][]{netzer87}.
This angular dependence of disk emission is adopted in 
SKIRTOR, the UV-to-IR AGN module adopted in \xcig\ 
(see \S\ref{sec:skirtor}).
The \xray\ emission should likely be less anisotropic than the
UV/optical emission, because the \hbox{X-rays} originate
from re-processed UV/optical photons via inverse Compton 
scattering.
However, the exact relation between \xray\ flux and viewing 
angle depends on model details, such as corona shape and 
opacity, which are poorly known \citep[e.g.][]{liu14, xu_y15}.
For simplicity, we assume that the \xray\ emission is isotropic.

Our assumption of anisotropic UV/optical emission and 
isotropic \xray\ emission leads to a dependence of 
$\ox$-$\luv$ on viewing angle.
We further assume that the observed $\ox$-$\luv$ relation 
for type~1 AGNs reflects the intrinsic $\ox$-$\luv$ relation 
for all AGNs at a ``typical'' viewing angle of 
$\theta=30^\circ$.
This value approximates the probability-weighted viewing angle 
for type~1 AGNs, i.e.
\begin{equation}
\label{eq:theta_type1}
    \theta \approx 
    \frac{\int_0^{90^\circ-\Delta} \theta \sin\theta d\theta }
         {\int_0^{90^\circ-\Delta} \sin\theta d\theta } 
         \approx 30^\circ,
\end{equation}
where $\Delta$ denotes the angle between the equatorial plane 
and edge of the torus, i.e., half opening angle.
The typical value is $\Delta \approx 40^\circ$ from observations 
\citep[e.g.][]{stalevski16}.
Although $\Delta$ is a free parameter in \xcig, we do not 
recommend the user choose other values than $40^\circ$,  
as this value is favored by observations and is 
consistently adopted throughout the build-up of the $\xcig$ code.
The weight $\sin\theta$ is proportional to the probability 
for the viewing angle being $\theta$. 

We note that our SED fitting results (\S\ref{sec:test} and 
\S\ref{sec:uplim}) are not sensitive to the assumed typical 
$\theta$, and will not change significantly if adjusting
$\theta$ within the range of $\approx 10^\circ$--$50^\circ$.
In the \xcig\ output (Appendix~\ref{sec:in_and_out}), the $\ox$ 
and $\luv$ always refer to the value at $\theta=30^\circ$, 
regardless of the actual viewing angle in the model.  
This $\ox$ and $\luv$ design is to reflect AGN essential
properties, independent of the viewing angle. 
By changing the integral ranges in Eq.~\ref{eq:theta_type1} 
to $(90^\circ-\Delta, 90^\circ)$, we can derive the 
probability-weighted $\theta$ for type~2 AGNs, i.e. 
$\theta \approx 70^\circ$.
These typical $\theta$ values (type~1: $\approx 30^\circ$, 
type~2:$\approx 70^\circ$) are used in our SED fitting 
(\S\ref{sec:test} and \S\ref{sec:uplim}).


\begin{figure}
\includegraphics[width=\columnwidth]{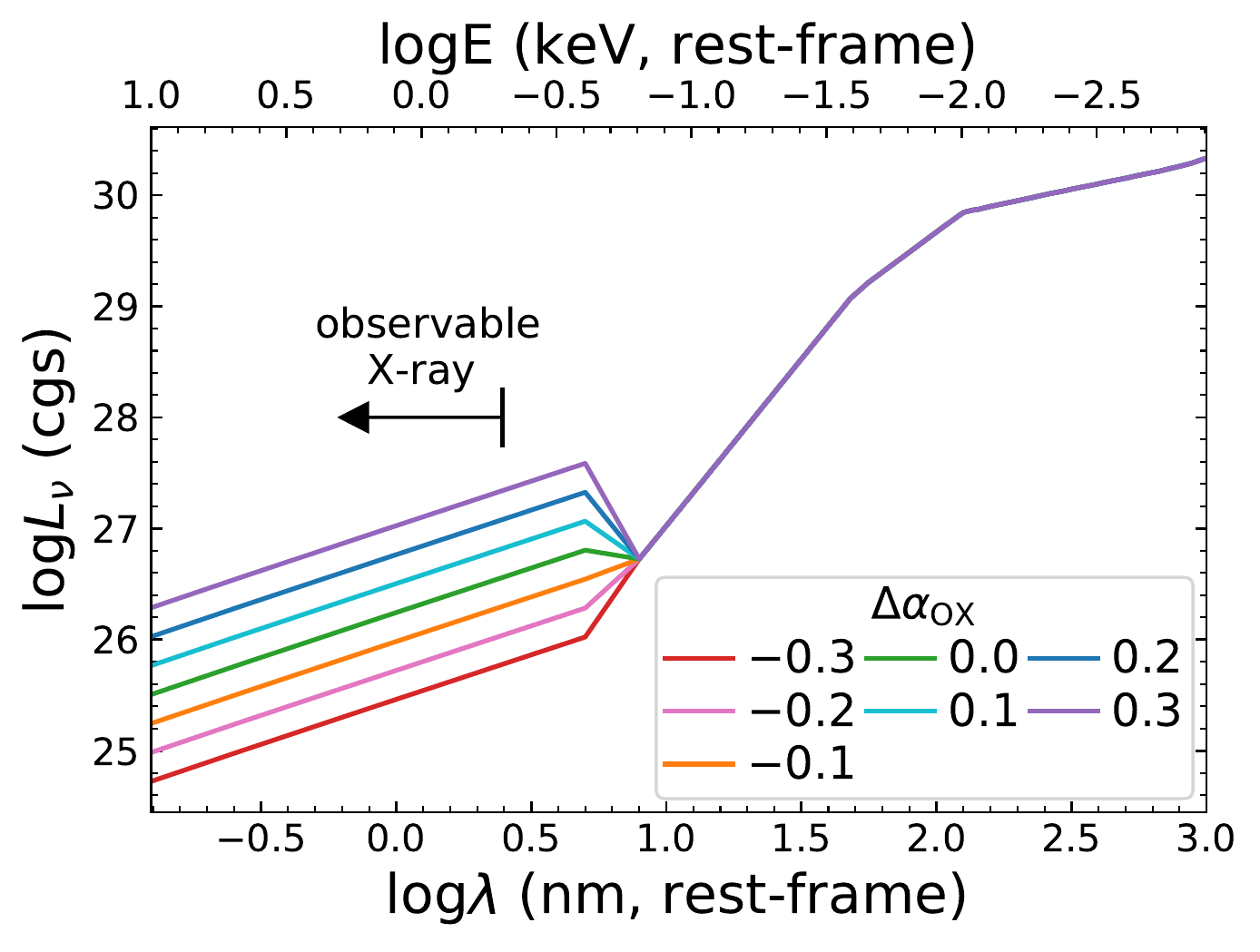}
\caption{The model SEDs for an unobscured AGN with 
$\log\luv = 30$ (cgs). 
Different colors indicate different $\ox$.
The ``breaks'' at $5$~nm are caused by the 
wavelength limit of the \xray\ module, which ends at 5~nm.
As explained in \S\ref{sec:ox}, such breaks are generally not 
problematic for practical purposes, as most \xray\ instruments only 
cover wavelengths $\lesssim 10^{0.4}$~nm ($\gtrsim 0.5$~keV, as marked).
}
\label{fig:sed_ox}
\end{figure}

\subsection{SKIRTOR}\label{sec:skirtor}
The previous CIGALE AGN model responsible for the UV-to-IR SED is 
from \cite{fritz06}. 
This model assumes that the dusty torus is a smooth structure.
However, more recent theoretical and observational works find
that the torus is mainly made of dusty clumps \citep[e.g.][]{nikutta09, 
ichikawa12, stalevski12, tanimoto19}.
SKIRTOR is a clumpy two-phase torus model \citep{stalevski12, stalevski16}, 
based on the 3D radiative-transfer code, SKIRT 
\citep{baes11, camps15}.\footnote{http://www.skirt.ugent.be/root/\_landing.html}
In SKIRTOR, most (mass fraction $=97\%$) of the dust is in the form of 
high-density clumps, while the rest is smoothly distributed.
In addition, SKIRTOR considers the anisotropy of the power source, 
AGN disk emission (see \S\ref{sec:ox}), while Fritz's model simply
assumes isotropic disk emission.
Therefore, we implement SKIRTOR within \xcig. 
We recommend using SKIRTOR as the UV-to-IR SED model 
of AGNs, although \xcig\ allows the user to choose between SKIRTOR 
and Fritz's model.

SKIRTOR adopts a disk SED that has a higher fraction of far-UV 
luminosity ($\lesssim 100$~nm) compared to observations 
(see \S3.2.1 of \hbox{\citealt{duras17}}).
Following \cite{duras17}, we update SKIRTOR with a new disk SED 
\citep{feltre12} that is supported by observations, i.e.
\begin{equation}
\label{eq:disk_sed}
\lambda L_\lambda \propto 
\begin{cases}
    \lambda^{2}   &\qquad 8 \leq \lambda < 50\ [\mathrm{nm}] \\
    \lambda^{0.8}  &\qquad 50 \leq \lambda < 125\ [\mathrm{nm}] \\
    \lambda^{-0.5} &\qquad 125 \leq \lambda < 10^4\ [\mathrm{nm}] \\
    \lambda^{-3}   &\qquad \lambda > 10^4\ [\mathrm{nm}].
\end{cases}
\end{equation}
We modify the disk SED with the following method.
We denote the old and new intrinsic disk SEDs as
$L^{\rm old}_{\lambda, \rm normed}$ and $L^{\rm new}_{\lambda, \rm normed}$, 
respectively, where the subscript ``normed'' indicates the total power 
of these SEDs has been normalized to unity.
Then the new observed disk SED component (which might be obscured) can 
be converted from the old one by multiplying by the factor, 
$L^{\rm new}_{\lambda, \rm normed}/L^{\rm old}_{\lambda, \rm normed}$.
The new scatter component can be obtained in the same way;
the dust reemitted component remains unchanged. 
The method above keeps energy balance.
This method is also described on the SKIRTOR official 
webpage.\footnote{https://sites.google.com/site/skirtorus/sed-library}

\subsection{Polar Dust}
\label{sec:polar_dust}

\subsubsection{The extinction of type~1 AGN}
\label{sec:type1_obsc}
In SKIRTOR (also in Fritz's model), the extinction of UV and optical 
radiation for type~1 AGN is assumed to be negligible.
{This assumption holds for most optically selected blue 
quasars. 
For example, \cite{richards03} found only $\approx 6\%$ 
of their SDSS quasars are extincted. 
However, the assumption might not be true for, e.g. \xray\ 
selected AGNs. 
For example, in the COSMOS AGN catalogs selected by \xmm\ 
\citep{bongiorno12}, the fraction of extincted sources ($E(B-V)\geq 0.1$) 
among broad-line AGNs is $\approx 40\%$.
}

\begin{figure*}
\includegraphics[width=\columnwidth]{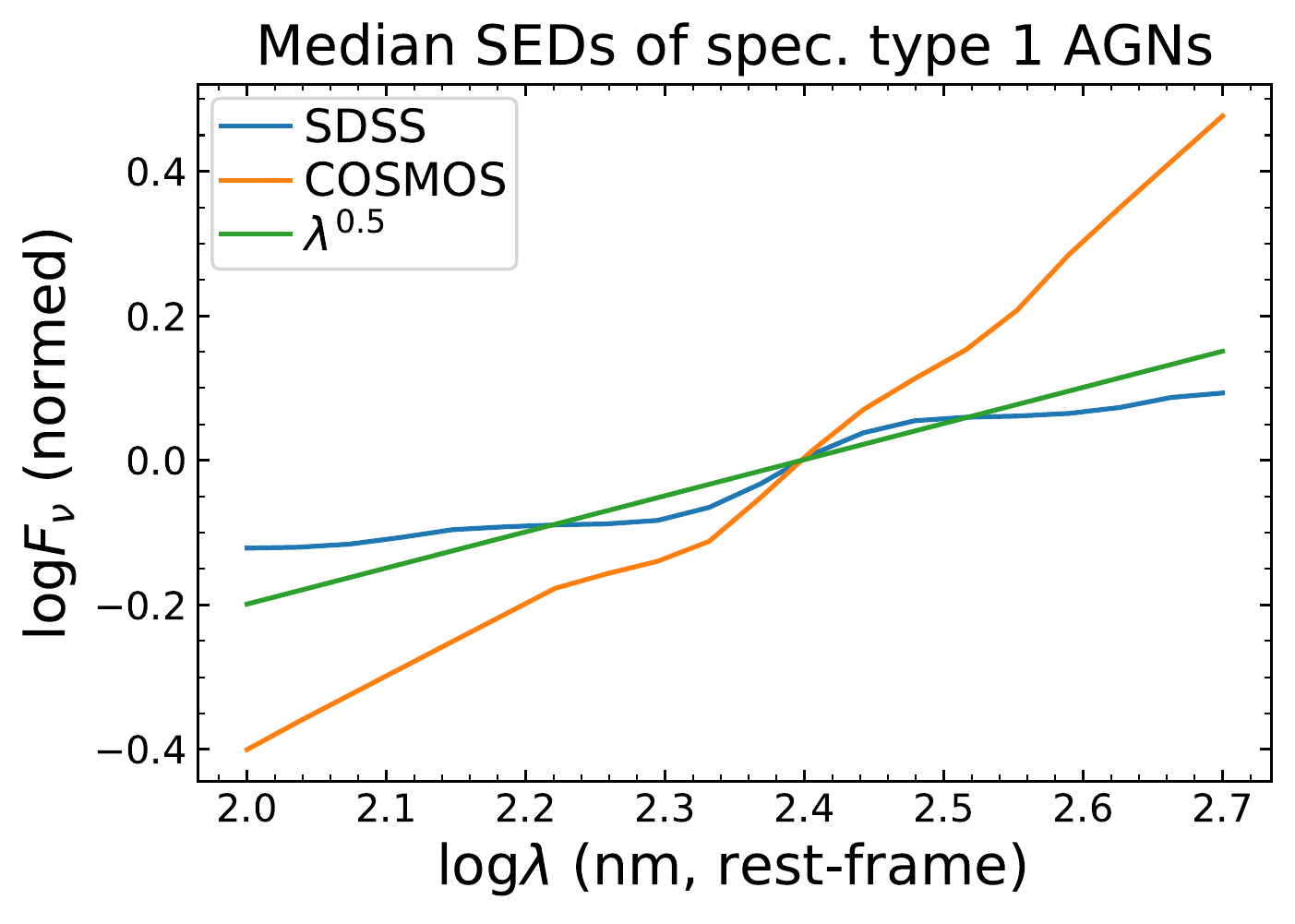}
\includegraphics[width=\columnwidth]{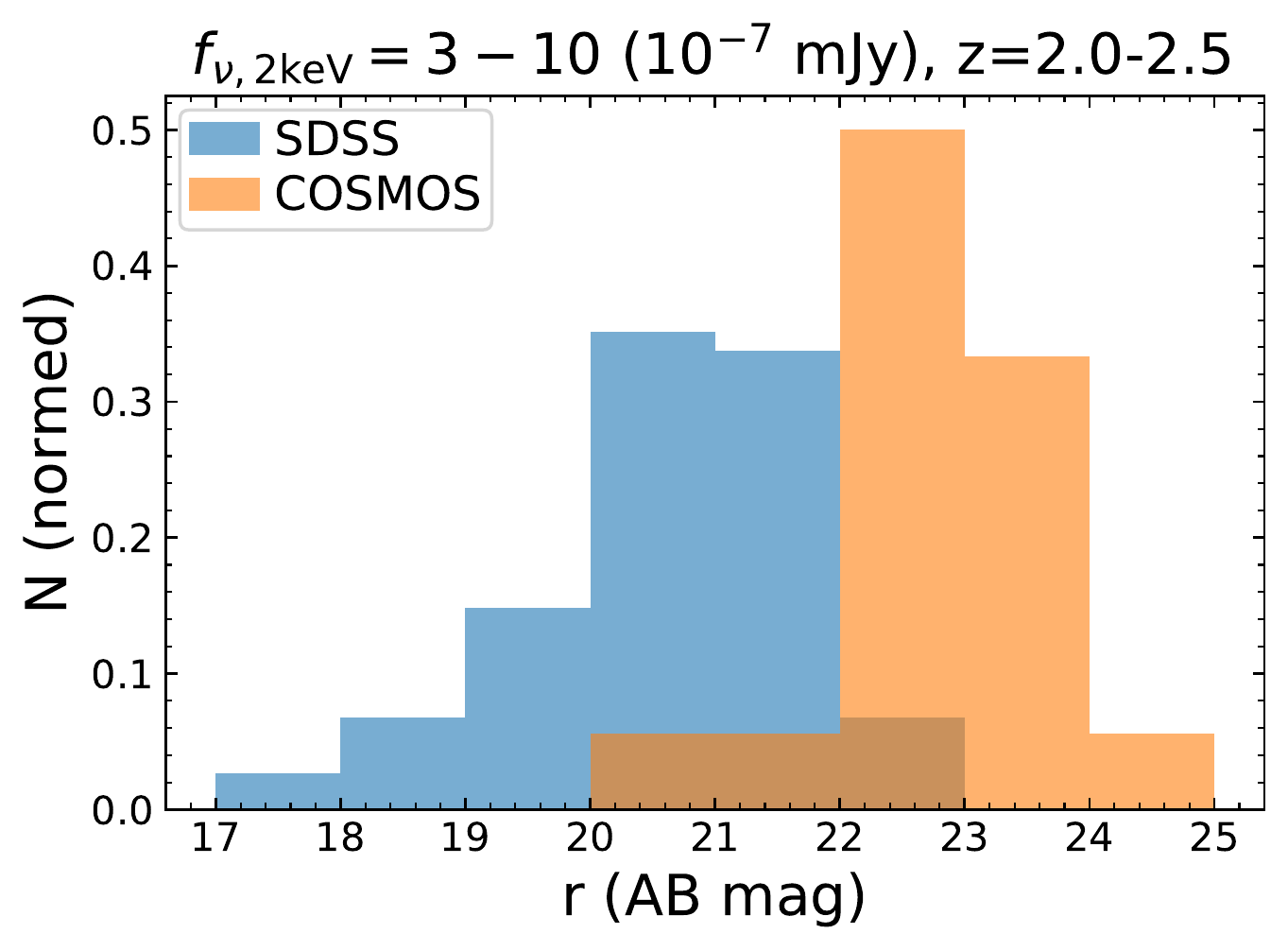}
\caption{\textit{Left}: {Median SEDs of spectroscopic 
type 1 AGNs in SDSS (blue, optically selected) and COSMOS 
(orange, \xray\ selected).} 
The green curve shows a typical unobscured quasar SED. 
All the SEDs are normalized at 250~nm.  
The SDSS SED is similar to the typical quasar SED.
The COSMOS SED is significantly redder than the SDSS SED,
indicating the presence of dust extinction and/or host-galaxy 
contribution for the COSMOS sources.
\textit{Right}:  
The observed $r$-band magnitude distributions for SDSS and 
COSMOS type~1 AGNs. 
Both AGN samples are in the bins of $z=2\text{--}2.5$ and 
$f_{\nu, \rm 2 keV} = 3\text{--}10\ (10^{-7}\ \rm mJy)$.
The SDSS AGNs are systematically brighter than the COSMOS AGNs, 
and this qualitative result also holds for other redshift/\xray\ 
flux bins.
{This result indicates that dust extinction should
be at least one of the causes for the red SED shapes of the 
COSMOS AGNs, although galaxy SED contributions might enhance 
the reddening.
This figure does not compare the \hbox{AKARI-NEP} sample 
in \S\ref{sec:akari}, because spectroscopic identifications of 
type~1 AGNs are not available for \hbox{AKARI-NEP} 
(\S\ref{sec:sample_akari}).}
}
\label{fig:med_sed}
\end{figure*}

To check the extinction of type~1 AGNs, we compare the median 
UV-optical SEDs of {spectroscopically classified} type~1 
AGNs in SDSS and COSMOS (see \S\ref{sec:test} for details).
{
These median SEDs are derived from the photometric data in 
\S\ref{sec:test}. 
For each source in a sample, we interpolate the observed photometry
to obtain $F_\nu$ as a function of observed-frame wavelength.
We then shift this interpolated SED to rest-frame wavelength and 
normalize $F_\nu$ at 250~nm. 
Finally, at each wavelength, we obtain the median $F_\nu$ of all 
the sources in the sample.
}
Fig.~\ref{fig:med_sed} (left) shows the results.
The SDSS median SED is similar to the typical unobscured quasar SED of 
$F_\nu \propto \lambda^{0.5}$ (see Eq.~\ref{eq:disk_sed}).
In contrast, the COSMOS median SED is significantly redder than 
$F_\nu \propto \lambda^{0.5}$ {\citep[e.g.][]{elvis12}}. 
{
We note that this difference in SED shape is observationally
driven by selection effects. 
The SDSS sample consists of optically selected, and is thus biased toward 
blue and optically bright objects. 
The COSMOS sample consists of \xray\ selected objects and does not suffer 
from significant bias in the UV/optical \citep[e.g.][]{brandt15}. 
Although driven by selection effects, 
Fig.~\ref{fig:med_sed} (left) at least indicates that reddened AGN SEDs 
indeed exist, and we discuss the physical cause of the SED 
reddening below.
}

{The red SED shape might be physically caused by 
the aforementioned dust extinction.
However, another potential physical cause is host-galaxy 
contribution to the SED. 
Since the UV/optical SEDs of galaxies are generally redder than 
those of unobscured AGNs (e.g. Fig.~3 of \hbox{\citealt{salvato09}}),
AGN-galaxy mixed SEDs tend to be redder than pure AGN SEDs. 
To investigate the cause of SED reddening, we can compare the 
magnitudes that sample rest-frame UV wavelengths for COSMOS and 
SDSS, as galaxy contributions to the photometry should be 
small at UV wavelengths.\footnote{{This statement 
breaks if the AGN host galaxies are highly star-forming in 
general. 
However, the AGN hosts tend to have normal levels of star-formation
activity, as shown by previous studies 
\citep[e.g.][]{harrison12, stanley15}.}}
In Fig.~\ref{fig:med_sed} (right), we show the $r$-band 
(rest-frame $\lesssim 2000\ \ang$) magnitude distributions 
at $z=2-2.5$ and $f_{\nu, \rm 2 keV} = 3\text{--}10$\ ($10^{-7}\ \rm mJy$,
the results are similar for other redshift/\xray\ flux bins). 
The control of redshift and \xray\ flux is to force the 
compared samples to have similar \xray\ luminosities and thereby 
bolometric luminosities (assuming the \xray\ bolometric correction
factors are similar for the sources).
The COSMOS AGNs are systematically fainter than the SDSS AGNs in 
UV/optical. 
Therefore, Fig.~\ref{fig:med_sed} (right) indicates that,
at a given AGN bolometric luminosity, the rest-frame UV AGN 
luminosities of COSMOS are typically lower than those of SDSS,
supporting the existence of dust extinction.
We conclude that dust extinction is at least one of the physical 
causes of the SED reddening (Fig.~\ref{fig:med_sed} left), 
although galaxy SED contributions might enhance the reddening
\citep[e.g.][]{bongiorno12}.  
}

\subsubsection{The polar-dust model}
\label{sec:polar_mod}
\begin{figure}
\includegraphics[width=\columnwidth]{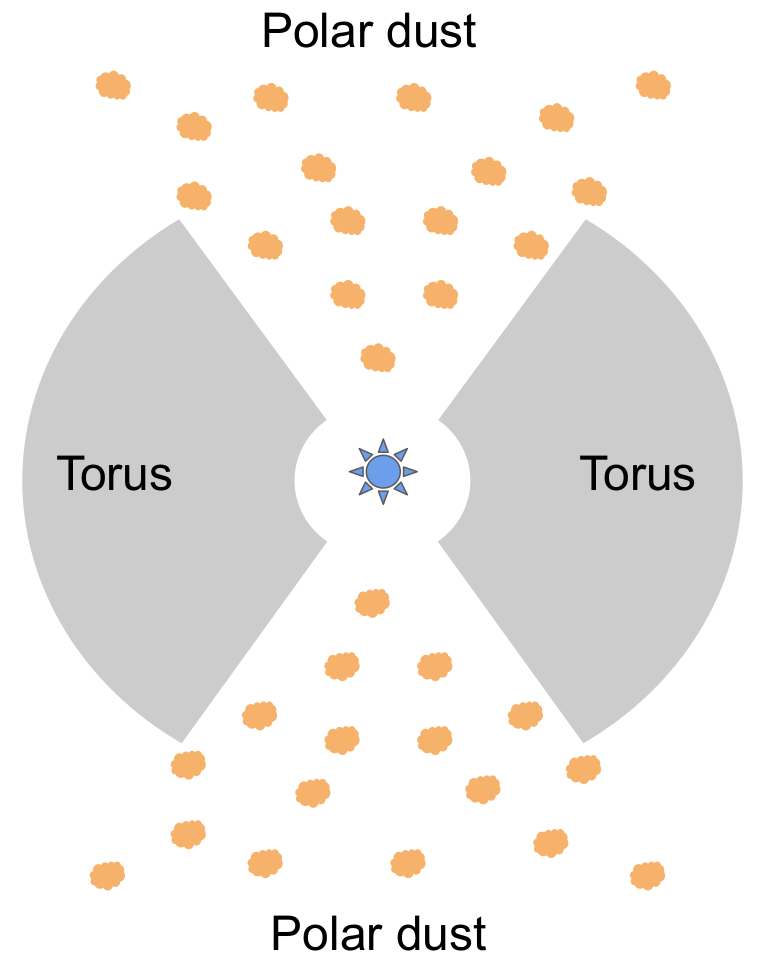}
\caption{Schematic plot (at the meridional plane) of the AGN 
obscuration system adopted by \xcig\ (not to scale).
The model in original CIGALE only includes torus obscuration. 
We add the obscuration of polar dust to account for type~1 
AGN extinction.
}
\label{fig:polar_dust}
\end{figure}

From \S\ref{sec:type1_obsc}, it is necessary to account for dust extinction 
of type~1 AGNs in \xcig. 
The geometry of the obscuring materials is sketched in
Fig.~\ref{fig:polar_dust}, where the materials responsible for type~1
AGN obscuration are called ``polar dust'' \citep[e.g.][]{lyu18}.
The existence of polar dust has been proved by high-resolution mid-IR 
(MIR) imaging of local Seyfert galaxies (e.g. 
\hbox{\citealt{lopez_gonzaga14}}; \hbox{\citealt{stalevski17, stalevski19}}; 
\hbox{\citealt{asmus19}}).
However, the physical properties of the polar dust could be complicated 
and vary for different objects.
For example, it might be close to the dust-sublimation radius 
($\sim$~pc scale; e.g. \citealt{lyu18}) or on galactic scales 
($\sim$~kpc; e.g. \citealt{zou19}).

Considering these complexities, we do not build a grid of physical 
models and perform radiation-transfer simulations.
Instead, we employ several empirical extinction curves, including 
those from 
\citet[][nearby star-forming galaxies]{calzetti00}, 
\citet[][large dust grains]{gaskell04}, 
and
\citet[][Small Magellanic Cloud, SMC]{prevot84},
and the user can choose among these curves.
In our tests of \xcig\ (\S\ref{sec:test} and \S\ref{sec:uplim}), 
we adopt the SMC extinction curve, which is preferred from AGN 
observations (e.g. \hbox{\citealt{hopkins04b}}; \hbox{\citealt{salvato09}}; 
\hbox{\citealt{bongiorno12}}; but also see, e.g. 
\hbox{\citealt{gaskell04}}).
The extinction amplitude (parameterized as $E(B-V)$) is 
a free parameter set by the user, and setting $E(B-V)=0$ 
returns to the original torus.

\begin{figure}
\includegraphics[width=\columnwidth]{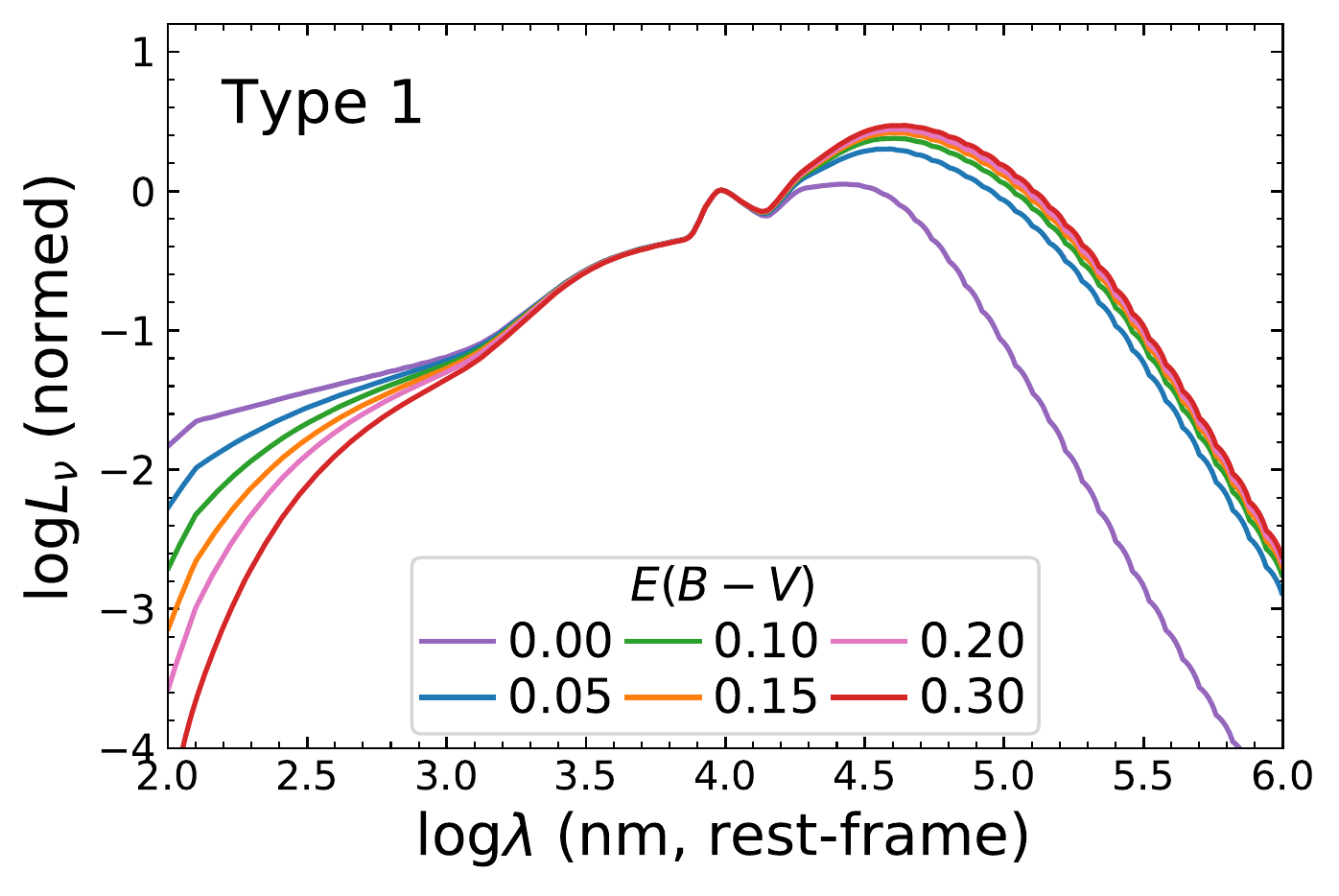}
\includegraphics[width=\columnwidth]{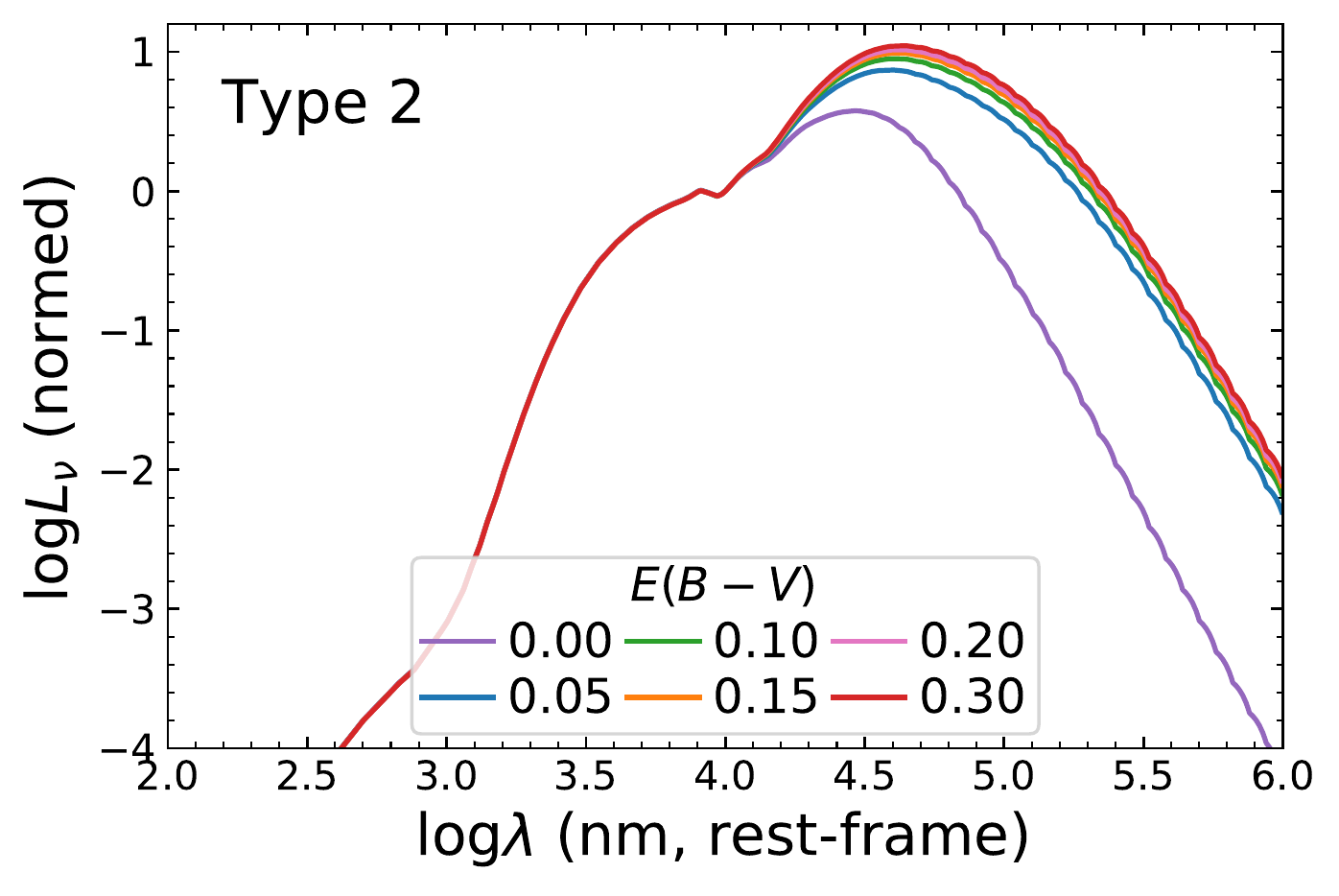}
\caption{Same format as in Fig.~\ref{fig:sed_ox}, but 
for AGNs with different polar-dust $E(B-V)$.
The top and bottom panels are for type~1 and type~2 AGNs, 
respectively.
The SEDs are normalized at $10\ \mu$m.
Except for $E(B-V)$, the other model parameters 
are the same (e.g. polar-dust temperature $=100$~K and 
emissivity $=1.6$) when generating these model SEDs.
}
\label{fig:sed_ebv}
\end{figure}

Since the scheme of \xcig\ maintains energy conservation, we need to 
implement dust emission to account for the radiative energy absorbed
by the dust.
We assume the dust reemission is isotropic. 
For simplicity, we adopt the ``grey body'' model
\citep[e.g.][]{casey12}, i.e.
\begin{equation}
\label{eq:pd_sed}
L_{\nu} (\lambda) \propto \frac{\left(1-\mathrm{e}^{-\left(\lambda_{0} / 
    \lambda\right)^{\beta}}\right)\left(\frac{c}{\lambda}\right)^{3}}{\mathrm{e}^{h 
    c / \lambda k T}-1},
\end{equation}
where $\lambda_0$ is fixed at 200~$\mu$m, emissivity ($\beta$) and 
temperature ($T$) are free parameters set by the user.
$L_{\nu}$ in Eq.~\ref{eq:pd_sed} is normalized so that total energy
is conserved, i.e.
\begin{equation}
\label{eq:em_ext}
L^{\rm emit}_{\rm total} = \int_{0}^{90^\circ-\Delta} 
        L^{\rm extinct}_{\rm total} (\theta) \sin\theta d\theta
\end{equation}
where $L_{\rm total}^{\rm emit}$ is the dust reemitted luminosity
(angle-independent) and $L^{\rm extinct}_{\rm total}$ is the 
luminosity loss caused by polar-dust extinction (angle-dependent).
Note that the integral on the right-hand-side of Eq.~\ref{eq:em_ext}
is to account for the fact that
{the polar dust only accounts for 
the obscuration in the polar directions while the 
polar-dust reemission is in all directions} 
(see Fig.~\ref{fig:polar_dust}; e.g. Eq.~\ref{eq:theta_type1}).
Fig.~\ref{fig:sed_ebv} shows the model SEDs for different extinction levels, 
where $T=100$~K, $\beta=1.6$, and $\Delta=40^\circ$.

{Our model above follows the AGN-unification scheme, i.e. 
AGN type is determined solely by the viewing angle, which is a 
free parameter in \xcig.
When the viewing angle is within the polar directions (type~1), the 
observed AGN disk emission suffers moderate (or none if $E(B-V)=0$) 
extinction from the polar dust.
When the viewing angle is within the equatorial directions (type~2), 
the observed AGN disk emission is strongly obscured by the torus.
If the AGN type is known (e.g. from spectroscopy), the user can 
limit the viewing angle to the polar or equatorial direction
(\S\ref{sec:sdss} and \S\ref{sec:cosmos}). 
Otherwise, the user can adopt multiple viewing angles including 
both polar and equatorial directions, and let \xcig\ choose freely 
between them (\S\ref{sec:akari}).
For example, if the user set ``viewing angles = 30$^\circ$, 
70$^\circ$; $E(B-V)=0.1$'', then CIGALE will build two model SEDs.
For the 30$^\circ$ model (type 1), the UV/optical SED is reddened 
by the $E(B-V)=0.1$ polar dust whose reemission also contributed 
to the IR SED. 
For the 70$^\circ$ model (type 2), the polar dust does not affect 
the UV/optical SED (already obscured by torus), but its reemission 
still contributes to the IR SED.
}

{Our polar-dust model above provides one possible scenario 
for the reddened type~1 AGNs, i.e. the viewing angle is small and 
the extinction is caused by dust along the polar directions.
An alternative scenario is that the line-of-sight (LOS)
intercepts the torus, but the extinction is only moderate by 
chance due to the inhomogeneity of torus.
However, this scenario has not been well investigated with physical 
torus models in the literature, to our knowledge. 
Therefore, we focus on the polar-dust model in the current version 
of \xcig, and future versions of \xcig\ may include this alternative 
scenario when its SED templates are available.
}


\section{Tests on X-ray detected sources}\label{sec:test}
In this section, we test \xcig\ with three samples of AGNs, i.e.
SDSS (\S\ref{sec:sdss}), COSMOS (\S\ref{sec:cosmos}), and 
\hbox{AKARI-NEP} (\S\ref{sec:akari}). 
The basic properties of these samples are summarized in 
Table~\ref{tab:samp_prop}.
These three samples have different characteristics.
The SDSS sample is optically bright type~1 quasars.
The COSMOS sample is \xray\ selected AGNs with broad multiwavelength 
coverage from $u$ to \herschel/PACS 160~$\mu$m.
This sample also has spectroscopic AGN classifications.
The \hbox{AKARI-NEP} sample is small but has excellent MIR 
observations from \akari.

\begin{table}
\centering
\caption{Sample properties}
\label{tab:samp_prop}
\begin{tabular}{cccccc} \hline\hline
Name & $N$ & $m_r$ & $f_{\rm 2\text{--}10\ keV}$ & Redshift & Type$_{\rm AGN}$ \\
\hline
     SDSS & 1986 &   19.2--20.9 & 2.2--10.5 & 0.6--1.9  &     1 \\
   COSMOS &  590 &   21.3--23.7 & 0.4-- 1.7 & 0.6--1.8  & 1 \& 2 \\
AKARI-NEP &   74 &   20.5--23.4 & 0.6-- 2.1 & 0.5--1.4  & 1 \& 2 \\
\hline
\end{tabular}
\begin{flushleft}
{\sc Note.} ---
(1) Survey name.
(2) Number of AGNs.
(3) $r$-band AB magnitude range (\hbox{20\%--80\%} percentile).
(4) \hbox{2--10 keV} \xray\ flux range (\hbox{20\%--80\%} percentile) 
    in units of $10^{-14}$~erg~s$^{-1}$~cm$^{-2}$.
(5) Redshift range (\hbox{20\%--80\%} percentile).
(6) Types of AGNs included in the survey.
\end{flushleft}
\end{table}

\subsection{SDSS}\label{sec:sdss}

\subsubsection{The sample and the models}\label{sec:sample_sdss}
The SDSS sample is optically selected from the DR14 quasar catalog
\citep{paris18}.
All the sources are spectroscopically confirmed type~1 AGNs.
In addition to the SDSS $ugriz$ bands, the \cite{paris18} catalog
provides \xray\ data from \xmm\ archival observations when 
available ({the 3XMM catalog}; 
\hbox{\citealt{rosen16}}).
We require the sources to be detected in the \hbox{2--12 keV} band 
at $>3\sigma$ significance levels. 
Here, the choice of the hard \xray\ band (\hbox{2--12 keV}) is to 
minimize the effects of \xray\ obscuration (see 
\S\ref{sec:set}).\footnote{Alternatively, one can perform \xray\ 
spectral fitting to obtain the absorption-corrected \xray\ fluxes.
However, extracting and analyzing the \xray\ spectra from 
the public \xmm\ archival data are beyond the scope of this work.}
{We do not include the IR photometry compiled in the DR14
catalog, because our main goal here is to test \xcig\ on the 
simple cases, i.e. the quasar-dominated SED. 
In the X-ray to optical wavelengths, the AGN component is dominant; 
but in the IR wavelengths, the galaxy component may be 
non-negligible.
We discuss the cases of AGN-galaxy mixed SEDs in \S\ref{sec:cosmos}
and \S\ref{sec:akari}.
}
We require the sources to have Galactic 
extinctions estimated in the DR14 quasar catalog, because we need
to correct for Galactic extinction before providing the photometry
to \xcig.
These criteria lead to a final sample of 1986 AGNs 
(Table~\ref{tab:samp_prop}).

For the SDSS sample, we can neglect the galaxy SED component, because
the sources are optically bright quasars which often dominate the 
observed UV/optical SEDs.
The AGN-dominant {(\xray\ to IR)} models in \xcig\ can be 
achieved by setting $\fracA$ to a value close to unity 
(e.g. 0.999).\footnote{Due to a technical reason, this value cannot 
equal to 1.}
The adopted AGN model parameters are listed in Table~\ref{tab:par_sdss}.
{The only free parameter in our fitting is polar-dust $E(B-V)$, 
which affects the UV/optical SED shape.}
We further justify that it is necessary to have $E(B-V)$ as a free 
parameter in \S\ref{sec:res_sdss}.
Other SKIRTOR parameters are fixed, because they only affect the IR 
SED shape where there is no band coverage for the SDSS sample
(see \S\ref{sec:res_akari} for the assessment of these parameters).

For the \xray\ module, we adopt $\Gamma=1.8$ 
{for AGN (the dominant component in \hbox{X-rays})}, 
the typical intrinsic photon index constrained by observations 
(\S\ref{sec:xray_sed}).
{Adopting other AGN $\Gamma$ values (e.g. 1.4 or 2.0) do not 
affect our fitting results significantly.
}
{Our adopted $\Gamma=1.8$ is slightly different from that assumed 
in the 3XMM catalog ($\Gamma=1.7$; \hbox{\citealt{rosen16}}).
Therefore, we scale the \hbox{2--12 keV} fluxes by a factor of 0.96 to correct
the effects of different $\Gamma$, and this correction factor is 
obtained using the PIMMS website.\footnote{http://cxc.harvard.edu/toolkit/pimms.jsp}
For the LMXB and HMXB components, we set $\Gamma=1.56$ and 2.0, respectively
(see \S\ref{sec:xray_sed}).
}
We adopt $|\Delta \ox|_{\rm max}=0.2$, and this $|\Delta \ox|_{\rm max}$ value 
is $\approx 2\sigma$ scatter of the $\ox$-$\luv$ relation (\S\ref{sec:ox}).
Note that although the \xray\ module has both parameters fixed, 
\xcig\ internally calculates 9 models of different $\ox$ values 
and selects $|\Delta \ox| \leq |\Delta \ox|_{\rm max}$ (see 
\S\ref{sec:ox}).


\begin{table*}
\centering
\caption{SDSS fitting parameters}
\label{tab:par_sdss}
\begin{tabular}{lll} \hline\hline
Module & Parameter & Values \\
\hline
\multirow{11}{*}{\shortstack[l]{AGN (UV-to-IR): \\ SKIRTOR}} 
    & Torus optical depth at 9.7 microns $\tau_{9.7}$ & 7.0 \\
    & Torus density radial parameter $p$ 
        ($\rho \propto r^{-p} \mathrm{e}^{-q|\cos (\theta)|}$) & 1.0 \\
    & Torus density angular parameter $q$ 
        ($\rho \propto r^{-p} \mathrm{e}^{-q|\cos (\theta)|}$) & 1.0 \\
    & Angle between the equatorial plane and edge of the torus $\Delta$ & 40$^\circ$ \\
    & Ratio of the maximum to minimum radii of the torus & 20 \\
    & Viewing angle $\theta$ (face on: $\theta=0^\circ$, edge on: 
                              $\theta=90^\circ$) & 30$^\circ$ \\
    & AGN fraction in total IR luminosity $\fracA$ & 0.999 \\
    & Extinction law of polar dust & SMC \\
    & $E(B-V)$ of polar dust & 0, 0.05, 0.1, 0.15, 0.2, 0.3 \\
    & Temperature of polar dust (K) & 100 \\
    & Emissivity of polar dust & 1.6 \\
\hline
\multirow{4}{*}{\shortstack[l]{X-ray: \\ This work}} 
    & AGN photon index & 1.8 \\ 
    & Maximum deviation from the $\ox$-$\luv$ relation 
    $|\Delta \ox|_{\rm max}$ & 0.2 \\
    & {LMXB photon index} & 1.56 \\ 
    & {HMXB photon index} & 2.0 \\ 
\hline
\end{tabular}
\end{table*}

\subsubsection{Fitting results}\label{sec:res_sdss}
\begin{figure}
\includegraphics[width=\columnwidth]{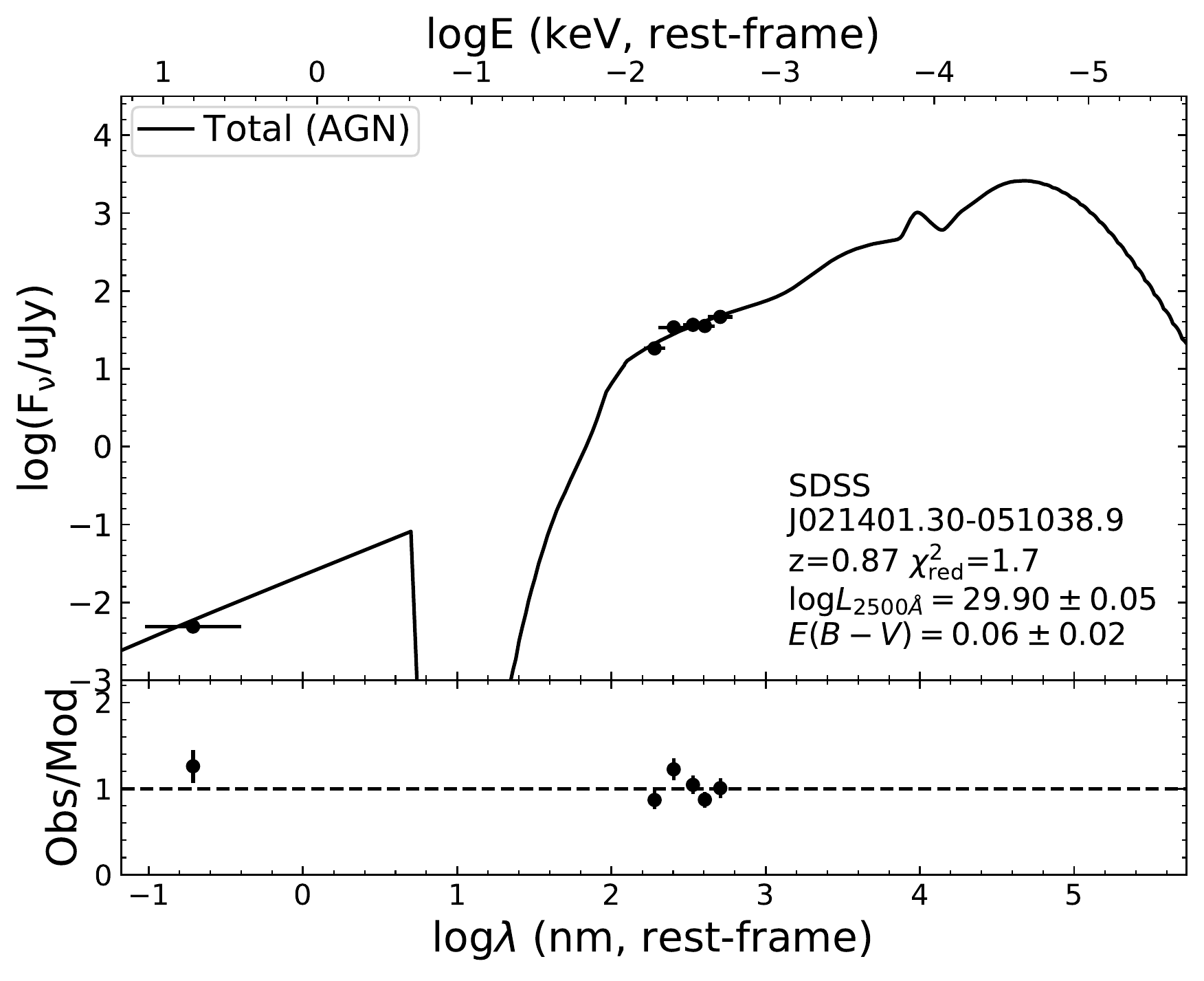}
\caption{A random example of fitted SED from SDSS.
The fitting residuals (observed flux/model flux) are shown
at the bottom.
{Some source properties are labelled on the plot.}
Since the fitting only uses the AGN component, the total SED
is actually the same as the AGN SED.
}
\label{fig:example_sed_sdss}
\end{figure}

\begin{figure}
\includegraphics[width=\columnwidth]{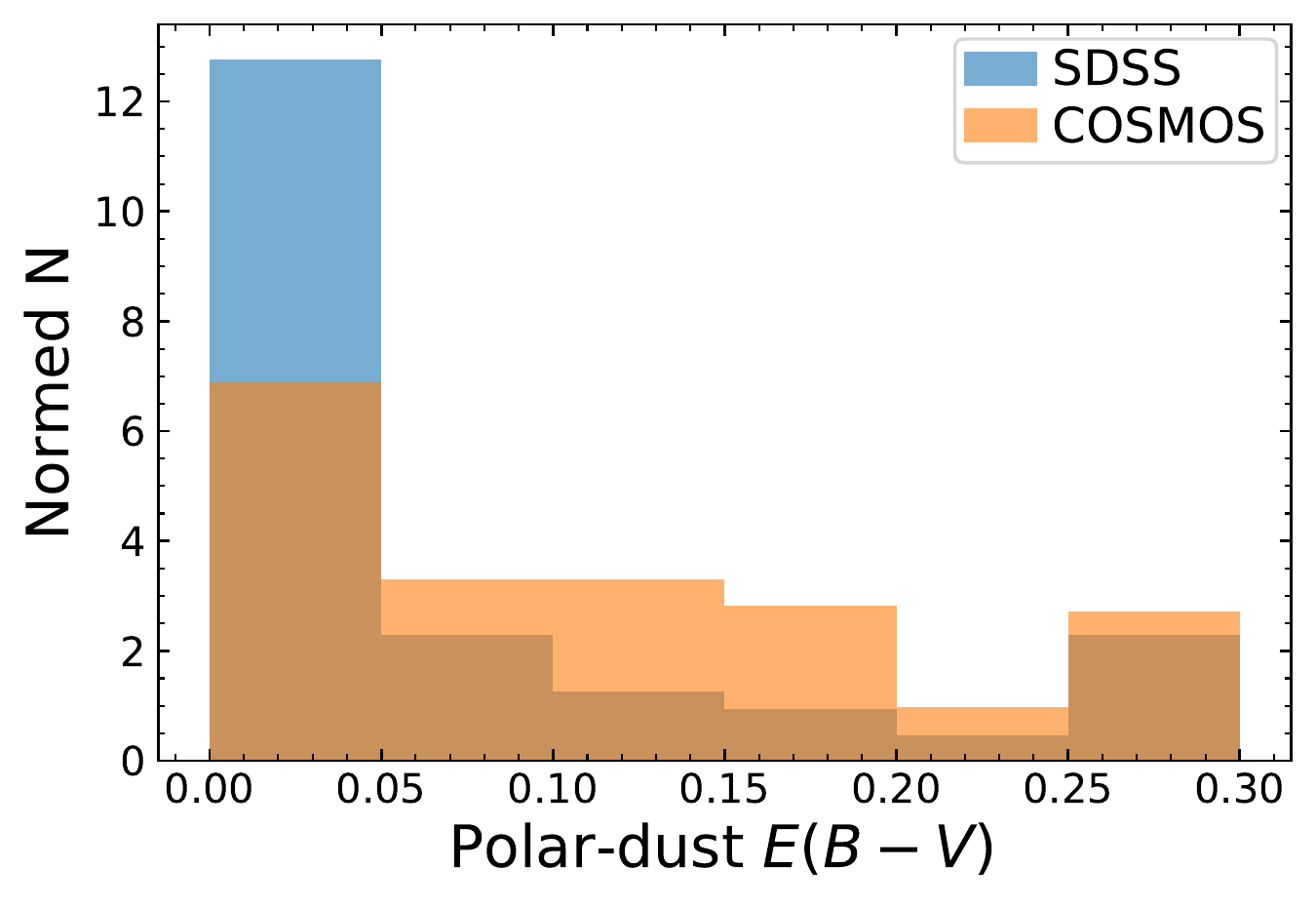}
\caption{The distribution of AGN polar-dust $E(B-V)$ 
({Bayesian-like} estimate) from \xcig\ fitting for SDSS (blue)
and COSMOS (orange) spectroscopic type~1 AGNs.
The histogram is normalized such that the integral 
is unity.
The COSMOS sources tend to have higher $E(B-V)$ than 
the SDSS sources.
}
\label{fig:ebv_hist}
\end{figure}

We run \xcig\ with the model settings in \S\ref{sec:sample_sdss}
for the SDSS sample.
The median reduced $\chi^2$ ($\redchi$) and degrees of freedom 
(dof) are 1.4 and 5, respectively. 
These $\redchi$ and dof values correspond to a $p$-value 
of 23\%, well above the conventional 2$\sigma$ (5\%) or
3$\sigma$ (0.3\%) values.
This result indicates that \xcig\ is able to 
{model} the observed photometry of the SDSS quasars.
Fig.~\ref{fig:example_sed_sdss} shows {a random example} 
of the SED fitting.
Fig.~\ref{fig:ebv_hist} displays the $E(B-V)$ distribution from 
the fitting. 
As expected (see \S\ref{sec:type1_obsc}), most (75\%) SDSS AGNs 
have weak or no extinction with $E(B-V)\leq 0.1$.

\begin{figure}
\includegraphics[width=\columnwidth]{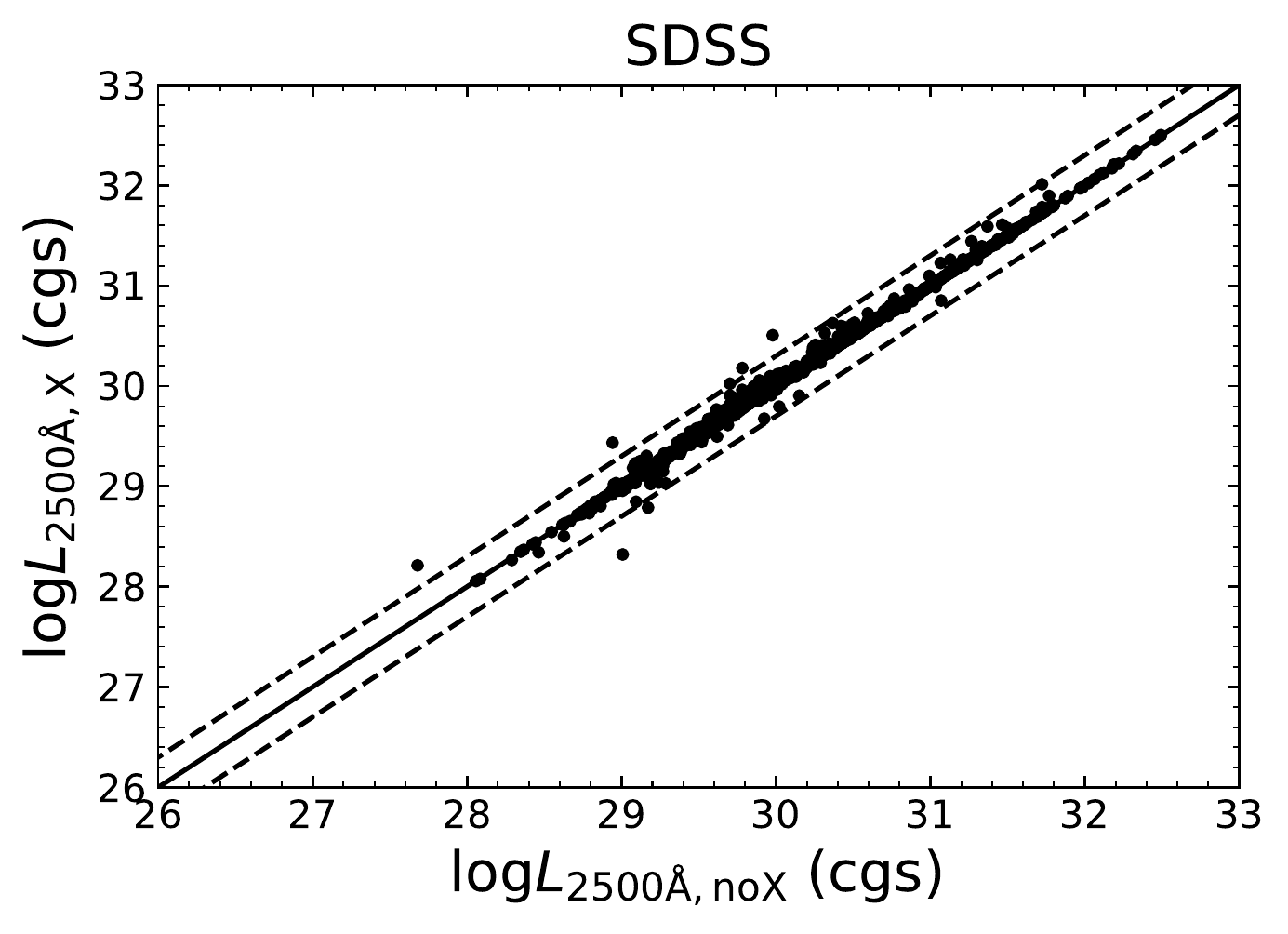}
\includegraphics[width=\columnwidth]{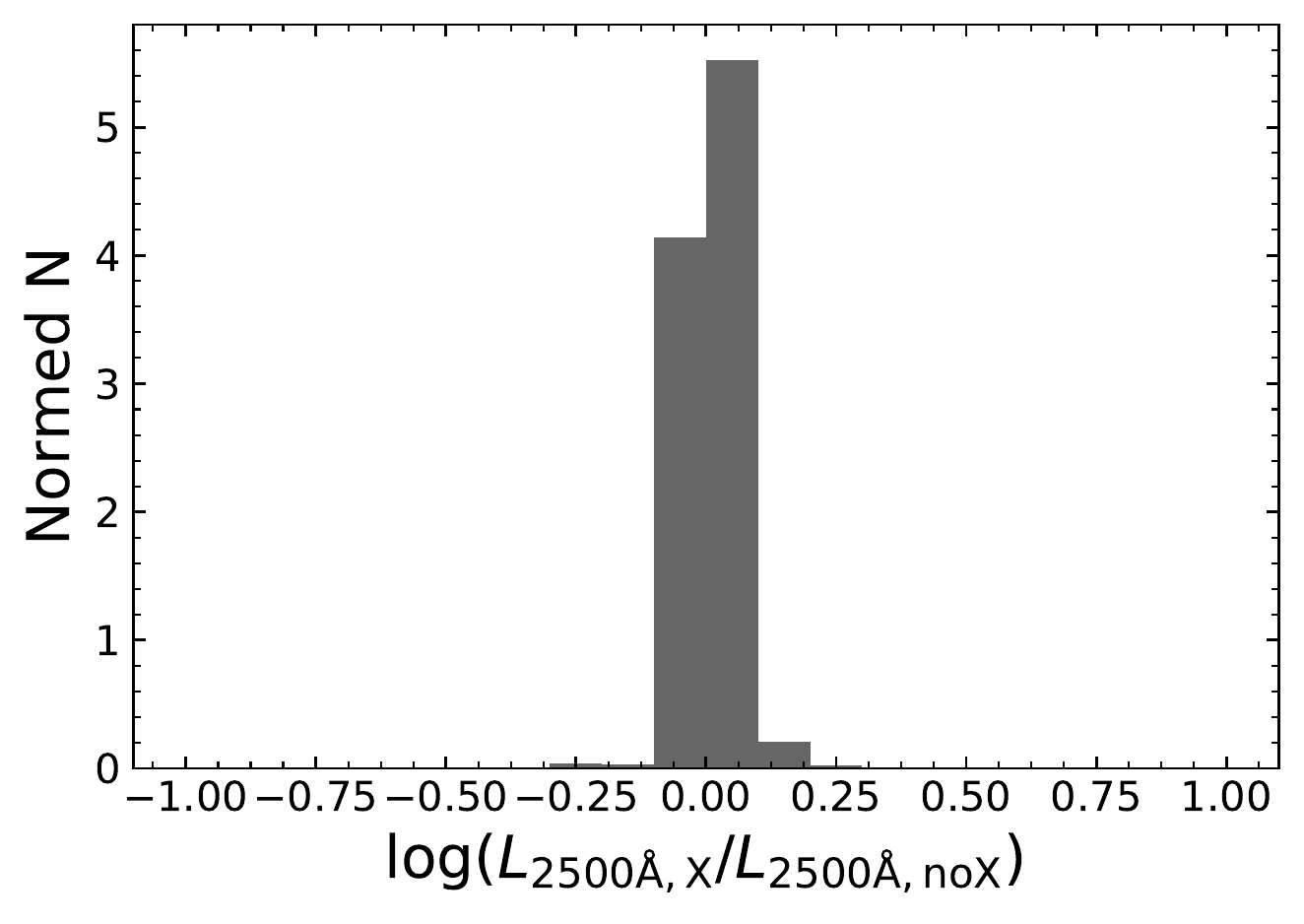}
\caption{Top: Comparison of AGN intrinsic $\luv$ between the 
fitting with \xray\ vs.\ without \xray\ for the SDSS sample.
Here, the $\luv$ values are {Bayesian-like} estimates of \xcig\ 
fitting.
The solid black lines indicate the 1:1 relation;
the dashed black lines indicate 0.3~dex deviation from 
the 1:1 relation.
Bottom: Distribution histogram of the $\luvx/\luvnox$ ratio for 
the SDSS sample.
The histogram is normalized such that the integral 
is unity.
}
\label{fig:Lbol_vs_Lbol_sdss}
\end{figure}

To evaluate the effects of the new \xray\ module, we re-run 
\xcig\ but without this module.
We compare the AGN intrinsic $\luv$ between the fitting with \xray\ 
($\luvx$) vs.\ without \xray\ ($\luvnox$) in Fig.~\ref{fig:Lbol_vs_Lbol_sdss}.
The $\luvx$ and $\luvnox$ are similar, and this similarity is 
as expected.
SDSS sources are mostly unobscured type~1 AGNs due to their selection 
method (\S\ref{sec:type1_obsc}), and thus the intrinsic AGN emission is 
directly observable at UV/optical wavelengths. 
Therefore, adding the \xray\ module does not significantly change $\luv$ 
estimation for SDSS sources in general.

In Table~\ref{tab:par_sdss}, the only free parameter is polar-dust
$E(B-V)$, because this parameter affects the UV/optical SED which is 
covered by the SDSS bands (\S\ref{sec:sample_sdss}).
In other words, we consider that $E(B-V)$ can be constrained by the 
photometric data.
The constrainability of a model parameter can be evaluated by 
the ``mock analysis'' of \xcig, which already exists in the previous
version of CIGALE (see \S4.3 of \hbox{\citealt{boquien19}} 
for details). 
Briefly, after fitting the observed data, \xcig\ simulates a mock 
catalog based on the best-fit model for each object.
The photometric uncertainties are considered when simulating the 
mock data.
\xcig\ then performs SED fitting to the mock catalog and obtains
{Bayesian-like} estimated values (and their errors) of the parameter.
By comparing these estimated values and those used to generate 
the mock catalog (i.e. the ``true'' values), 
one can assess whether the parameter can be reliably constrained.
{The mock analysis serves as a sanity check to assess 
whether a physical parameter can be retrieved in a self-consistent 
way.}
This mock analysis can be invoked by setting 
``mock\_flag$=$True'' in the \xcig\ configurations.

\begin{figure*}
\includegraphics[width=\columnwidth]{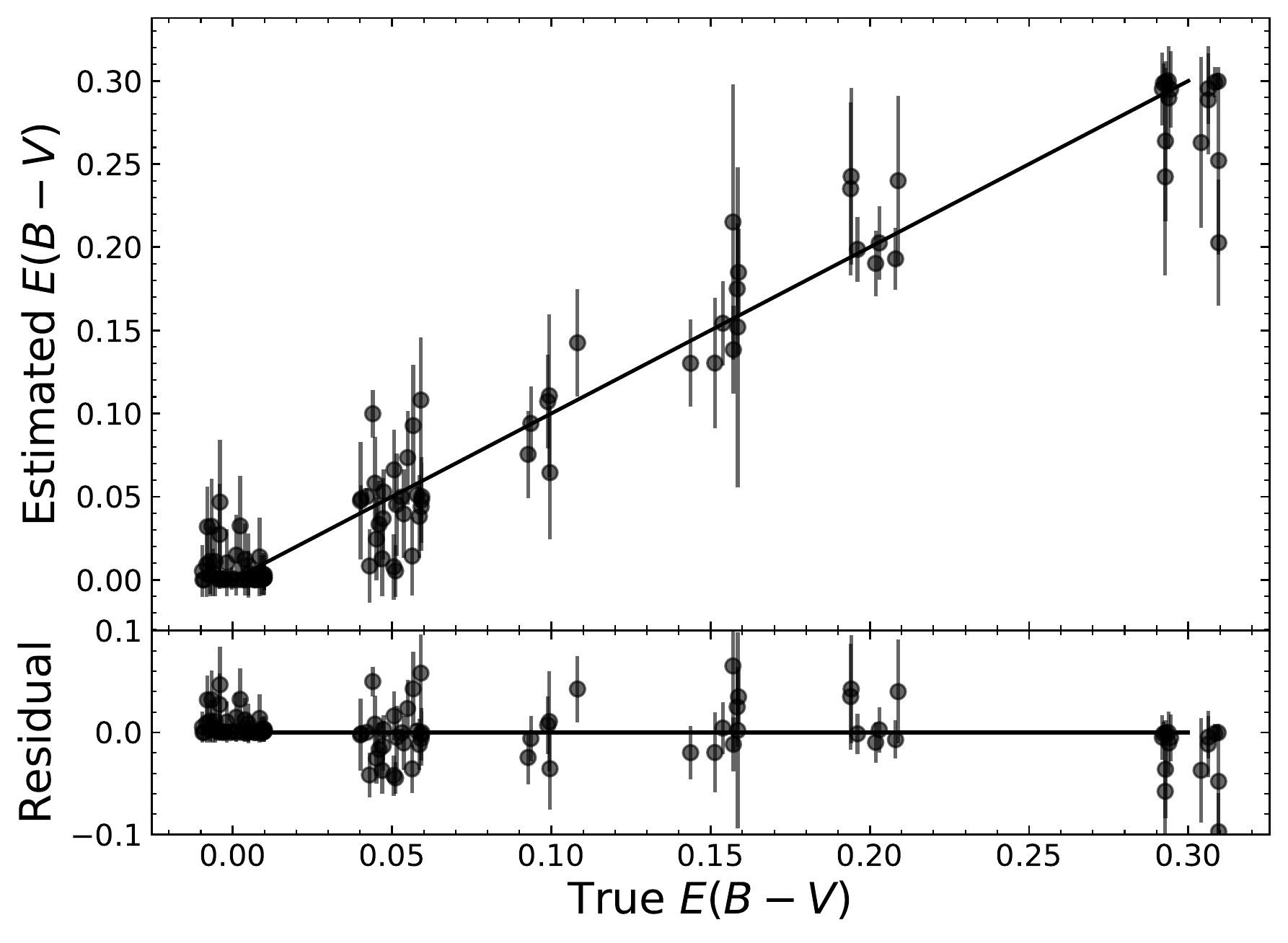}
\includegraphics[width=\columnwidth]{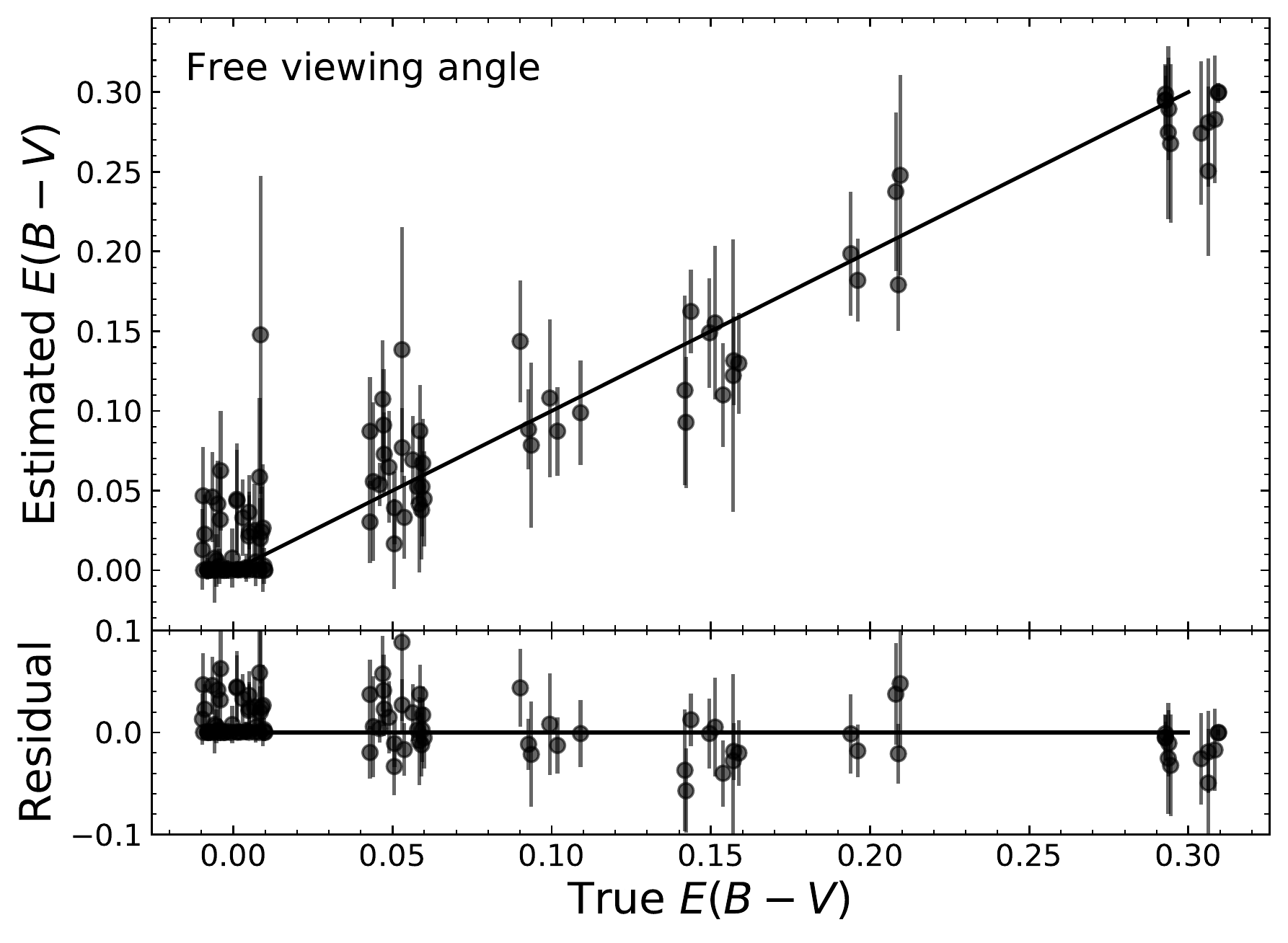}
\caption{{Left:} Estimated vs.\ true values in the mock analyses of 
the SDSS sample for polar-dust $E(B-V)$.
The black line indicates the 1:1 relation between estimated 
and true values.
For clarity, the data points are from 100 randomly selected 
sources in the SDSS sample.
The values on the $x$-axis are shifted slightly for display 
purposes only.
{The residuals (defined as estimated minus true) are 
displayed on the bottom panel, with the black line indicating
a zero residual.}
For $E(B-V)$, the estimated and true values are well correlated,
indicating that $E(B-V)$ model parameter can be effectively 
constrained by the data.
{Same format as left but from the fitting
with free viewing angle.}
}
\label{fig:mock_sdss}
\end{figure*}

\begin{figure}
\includegraphics[width=\columnwidth]{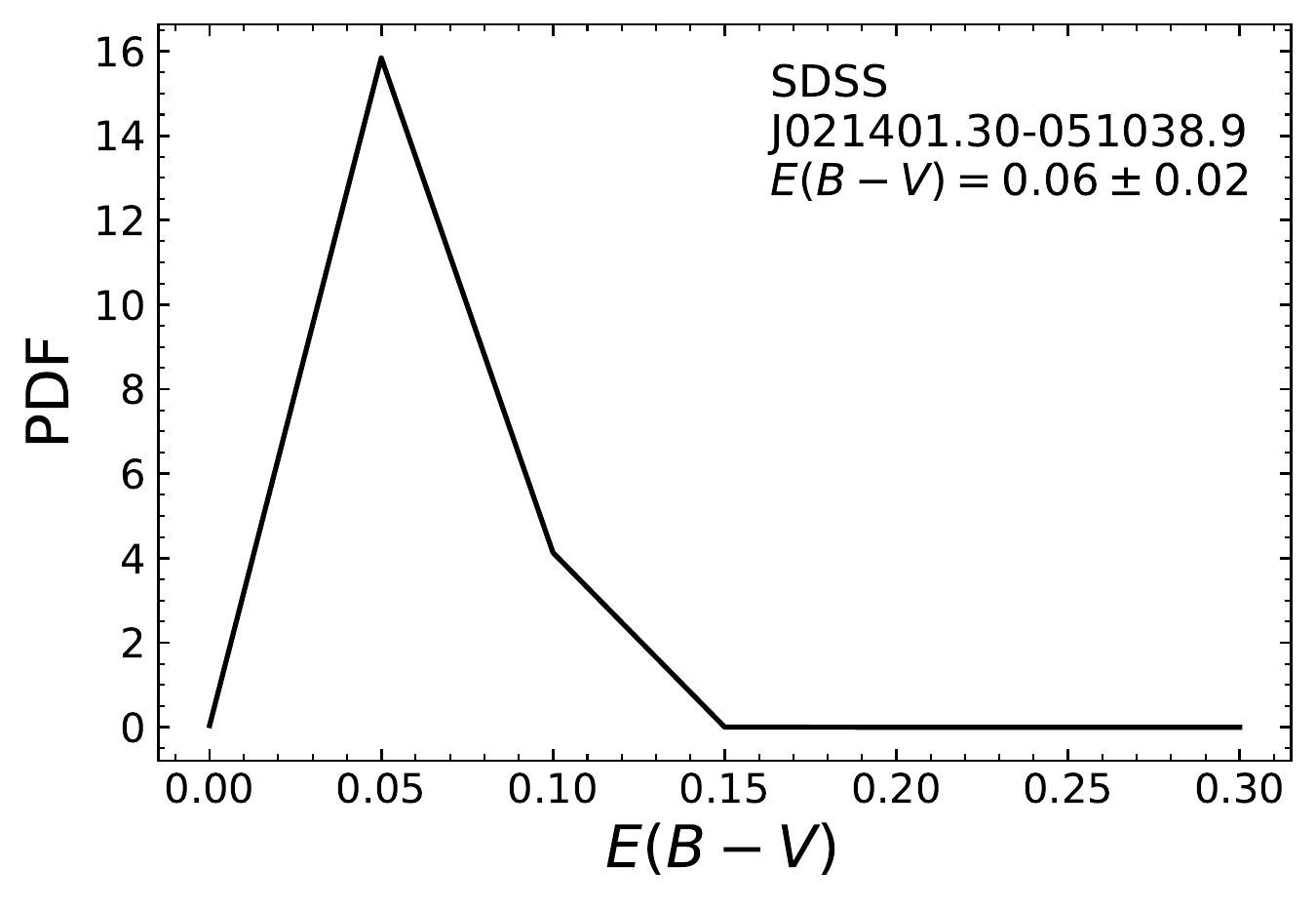}
\caption{{The $E(B-V)$ PDF for the example source in 
Fig.~\ref{fig:example_sed_sdss}. 
The PDF is normalized such that its integral is 
unity.
}
}
\label{fig:example_pdf_sdss}
\end{figure}

We run the mock analysis to test if polar-dust $E(B-V)$ can be 
constrained.
We compare the estimated and true values in Fig.~\ref{fig:mock_sdss} 
(left).
The estimated and true values are well correlated, indicating that
$E(B-V)$ can be {self-consistently constrained.
In Fig.~\ref{fig:example_pdf_sdss}, we show the PDF of $E(B-V)$
for the source in Fig.~\ref{fig:example_sed_sdss}. 
Fig.~\ref{fig:example_pdf_sdss} indicates that 
the $E(B-V)$ is indeed well constrained in the {Bayesian-like} 
analysis.
}

{In our fitting, aside from the model normalization 
(automatically determined by \xcig; see \S4.3 of \hbox{\citealt{boquien19}}), 
$E(B-V)$ is the only free model parameter (Table~\ref{tab:par_sdss}). 
We also test freeing other parameters such as viewing angle and 
torus optical depth, and the mock-analysis results of $E(B-V)$ are 
similar.
For example, Fig.~\ref{fig:mock_sdss} (right) shows the result
after setting the viewing angle to 0\text{--}90$^\circ$ with a step 
of 10$^\circ$ (i.e. all allowed values). 
The estimated and true values are still well correlated, indicating
that $E(B-V)$ and viewing angle are not strongly degenerate.
This non-degeneracy is understandable, because, in our polar-dust model, 
$E(B-V)$ is the only parameter responsible for modelling the observed 
UV/optical SED shapes of type~1 AGNs like the SDSS objects 
(\S\ref{sec:polar_dust}).
}


\subsection{COSMOS}\label{sec:cosmos}

\subsubsection{The sample and the models}\label{sec:sample_cosmos}
The COSMOS sample is \xray\ selected ($>3\sigma$, \hbox{2--10 keV} 
band) from the \hbox{COSMOS-Legacy} survey performed by \chandra\ 
\hbox{\citep{civano16}}.
{The \hbox{COSMOS-Legacy} catalog assumes $\Gamma=1.4$. 
Similarly as in \S\ref{sec:sample_sdss}, we apply a correction factor
of 0.87 (calculated with PIMMS) to the \hbox{2--10 keV} fluxes to 
make them consistent with our adopted $\Gamma=1.8$ for AGN.
}
\cite{marchesi16} matched these \xray\ sources with the optical/NIR
counterparts in the COSMOS2015 catalog \citep{laigle16} and compiled 
their spectroscopic information when available.
We select the sources with spectroscopic classifications of AGN 
types.
We adopt the photometric data in the COSMOS2015 catalog, including 
14 broad bands from $u$ to IRAC~8.0~$\mu$m.
In addition, when available, we also include photometric data from
\spitzer/MIPS (24~$\mu$m) and \herschel/PACS (100~$\mu$m and 160~$\mu$m), 
from the PEP survey \citep{lutz11}.
We adopt the redshift measurements from \cite{marchesi16}.
These redshifts are either secure spectroscopic redshifts 
or high-quality photometric redshifts. 
There are a total of 590 objects in COSMOS 
(Table~\ref{tab:samp_prop}).
Among these 590 objects, 206 and 384 are type~1 and type~2 AGNs, 
respectively. 

The \xcig\ model parameters for COSMOS are listed in Table~\ref{tab:par_cosmos}.
For the SKIRTOR and \xray\ modules, the parameter setting is the same as 
in \S\ref{sec:sample_sdss} except for $\fracA$ and the viewing angle.
Here, we allow $\fracA$ to vary among 0.01, 0.1--0.9 (step 0.1), and 0.99,
because, unlike in the case of SDSS, the AGN contribution to the observed SED 
may not be generally not dominant for the COSMOS sample, especially in the 
IR bands.
We set the viewing angle to 30$^\circ$ and 70$^\circ$ for the spectroscopic 
type~1 and type~2 AGNs, respectively.
These values are approximately the probability-weighted $\theta$ for type~1 and 
type~2 AGNs, respectively, given a torus of $\Delta = 40^\circ$ (see \ref{sec:ox}).

For the galaxy component, we adopt the model setting similar to that in 
\cite{ciesla15}.
Specifically, we adopt a delayed star-formation history (SFH), 
because it can characterize the SEDs of both early-type and late-type 
galaxies reliably \citep[e.g.][]{ciesla15, boquien19}.
Also, the delayed SFH only has relatively small parameter space (only two 
free parameters) and thereby high fitting efficiency.
We adopt a \cite{chabrier03} IMF with metallicity 
($Z$) fixed to the solar value of 0.02.
For the galactic dust attenuation, we adopt the dustatt\_calzleit module in 
\xcig\ (\hbox{\citealt{calzetti00}}; \hbox{\citealt{leitherer02}}).
The allowed $E(B-V)$ values for young stars are 0.1, 0.2, 0.3, 0.4, 0.5, 0.7, 
and 0.9.
The $E(B-V)$ ratio between the old and young stars is fixed to 0.44.  
The amplitude of the 217.5~nm UV bump on the extinction curve is set to 
0 (SMC) and 3 (Milky Way).
We adopt the \cite{dale14} model for galactic dust reemission.
There is only one free parameter in this model, i.e. the $\alpha$ slope 
in $dM_{\rm dust} \propto U^{-\alpha} dU$, where $M_{\rm dust}$ and $U$
are dust mass and radiation-field intensity, respectively.
The $\alpha$ values are set to 1.5, 2.0, and 2.5.

\begin{table*}
\centering
\caption{COSMOS and AKARI-NEP Fitting Parameters}
\label{tab:par_cosmos}
\begin{tabular}{lll} \hline\hline
Module & Parameter & Values \\
\hline
    \multirow{2}{*}{\shortstack[l]{Star formation history:\\ 
                                  delayed model, $\mathrm{SFR}\propto t \exp(-t/\tau)$ }} 
    & $e$-folding time, $\tau$ (Gyr) & 0.1, 0.5, 1, 5 \\
    & Stellar Age, $t$ (Gyr) & 0.5, 1, 3, 5, 7 \\ 
\hline
\multirow{2}{*}{\shortstack[l]{Simple stellar population:\\ \cite{bruzual03}}} 
    & Initial mass function & \cite{chabrier03} \\
    & Metallicity ($Z$) & 0.02 \\
\hline
    \multirow{2}{*}{\shortstack[l]{Galactic dust attenuation:\\ \cite{calzetti00} \& \cite{leitherer02} }} 
    & $E(B-V)$ of starlight for the young population  
        & 0.1, 0.2, 0.3, 0.4, 0.5, 0.7, 0.9 \\
    & $E(B-V)$ ratio between the old and young populations 
        & 0.44 \\
\hline
\multirow{1}{*}{\shortstack[l]{Galactic dust emission: \cite{dale14}}} 
    & $\alpha$ slope in $dM_{\rm dust} \propto U^{-\alpha} dU$
    & 1.5, 2.0, 2.5 \\
\hline
\multirow{11}{*}{\shortstack[l]{AGN (UV-to-IR): \\ SKIRTOR}} 
    & Torus optical depth at 9.7 microns $\tau_{9.7}$ & 7.0 \\
    & Torus density radial parameter $p$ 
        ($\rho \propto r^{-p} \mathrm{e}^{-q|\cos (\theta)|}$) & 1.0 \\
    & Torus density angular parameter $q$ 
        ($\rho \propto r^{-p} \mathrm{e}^{-q|\cos (\theta)|}$) & 1.0 \\
    & Angle between the equatorial plan and edge of the torus & 40$^\circ$ \\
    & Ratio of the maximum to minimum radii of the torus & 20 \\
    & Viewing angle $\theta$ (face on: $\theta=0^\circ$, edge on: 
                              $\theta=90^\circ$)$^a$ & 30$^\circ$ (type~1), 70$^\circ$ (type~2) \\
    & AGN fraction in total IR luminosity $\fracA$ & 0.01, 0.1--0.9 (step 0.1), 0.99 \\
    & Extinction law of polar dust & SMC \\
    & $E(B-V)$ of polar dust & 0, 0.05, 0.1, 0.15, 0.2, 0.3 \\
    & Temperature of polar dust (K) & 100 \\
    & Emissivity of polar dust & 1.6 \\
    \hline
\multirow{2}{*}{\shortstack[l]{X-ray: \\ This work}} 
    & AGN photon index $\Gamma$ & 1.8 \\ 
    & Maximum deviation from the $\ox$-$\luv$ relation & 0.2 \\
    & {LMXB photon index} & 1.56 \\ 
    & {HMXB photon index} & 2.0 \\ 
\hline
\end{tabular}
\begin{flushleft}
{\sc Note.} --- ($a$) For COSMOS, the viewing angles are set to 30$^\circ$ and 70$^\circ$ for 
                      the spectroscopic type~1 and type~2 AGNs, respectively. 
                      For \hbox{AKARI-NEP}, we allow \xcig\ to choose between 30$^\circ$ and 
                      70$^\circ$ for the entire sample, since spectroscopic classifications 
                      are not available.
\end{flushleft}
\end{table*}

\subsubsection{Fitting results}\label{sec:res_cosmos}
\begin{figure}
\includegraphics[width=\columnwidth]{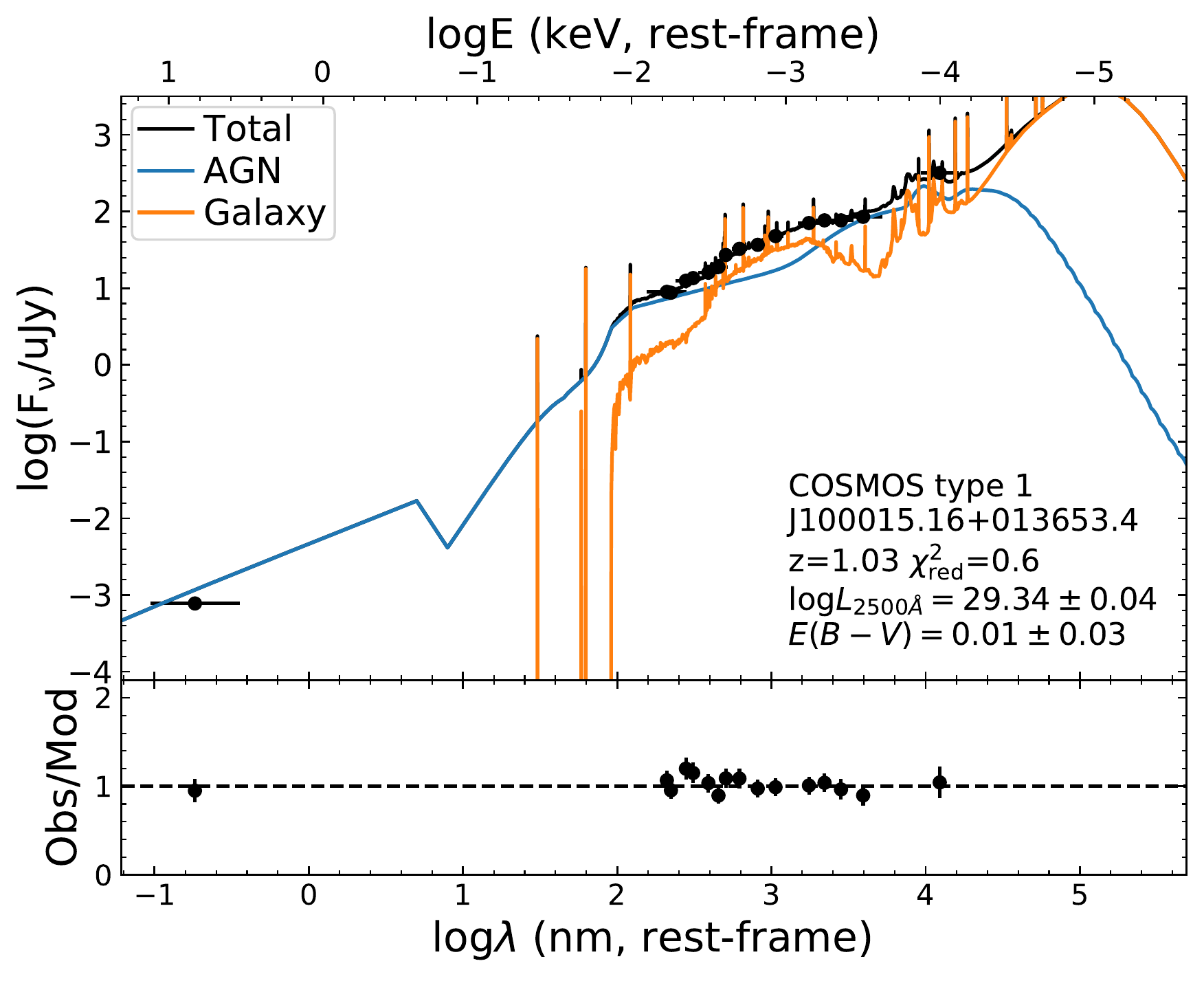}
\includegraphics[width=\columnwidth]{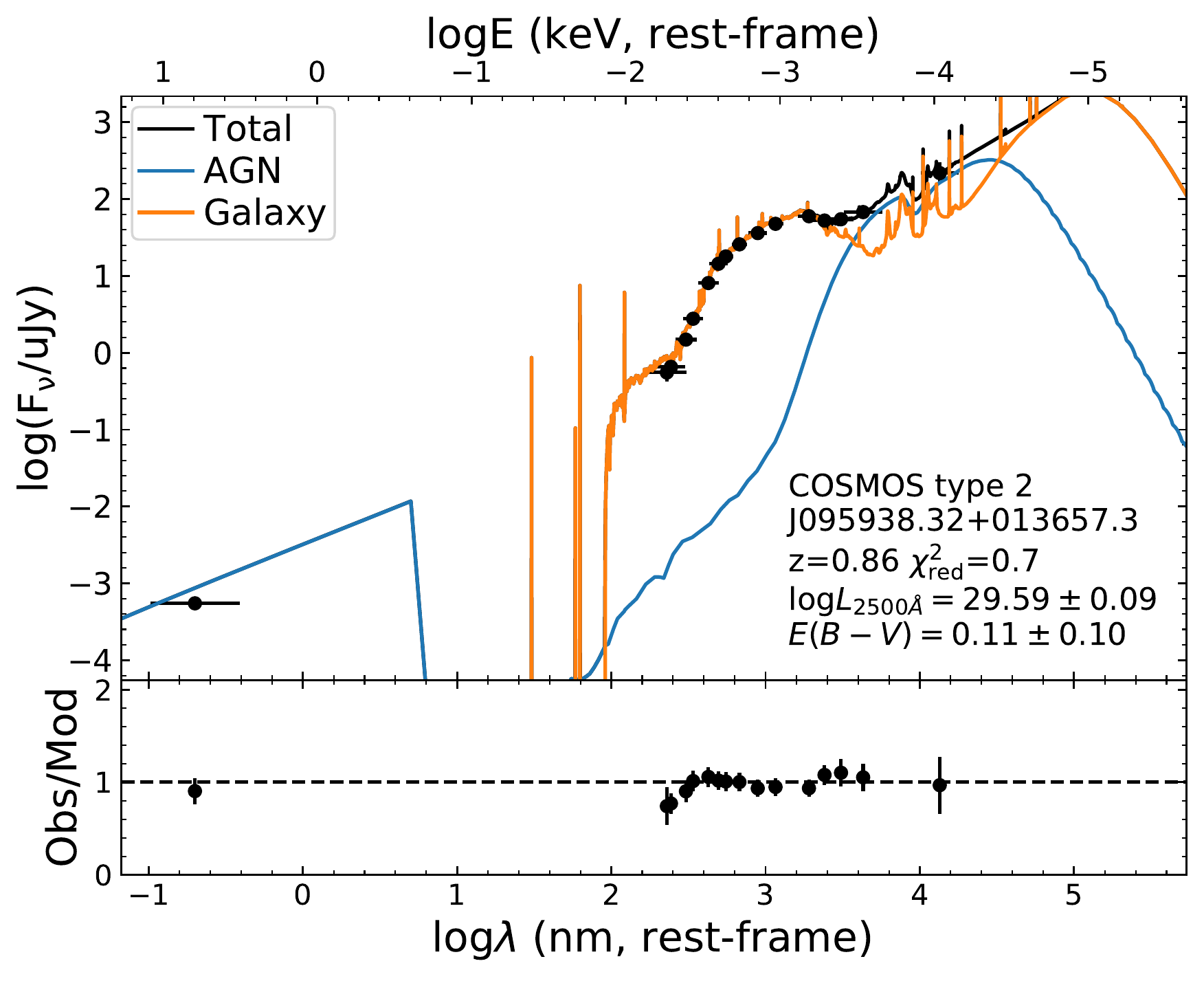}
\caption{Same format as in Fig.~\ref{fig:example_sed_sdss} 
but for COSMOS type~1 (top) and type~2 (bottom).
The orange and blue curves indicate galaxy and AGN 
model components.
}
\label{fig:example_sed_cosmos}
\end{figure}

We run \xcig\ with the parameter settings in \S\ref{sec:sample_cosmos}.
The median $\redchi$ values are 1.4 and 0.9, for type~1 and type~2
AGNs, respectively, while the median dof are 15 for both types.
These median $\redchi$ and dof corresponding to $p$-values of 0.12 and 
0.59.
These relatively large $p$-values for both type~1 and type~2 indicates
that our models (\S\ref{sec:code}) are able to {model} AGN SEDs 
of different types.
This result supports the AGN-unification scheme (\S\ref{sec:intro}), 
on which our models are based.
Fig.~\ref{fig:example_sed_cosmos} displays two examples of the SED 
fitting in COSMOS. 
In Fig.~\ref{fig:ebv_hist}, we compare the polar-dust $E(B-V)$ of type~1 
AGNs in COSMOS vs.\ SDSS.
The COSMOS type~1 AGNs tend to have higher $E(B-V)$ than SDSS type~1 AGNs, 
consistent with the diagnostic in \S\ref{sec:type1_obsc}.

\begin{figure}
\includegraphics[width=\columnwidth]{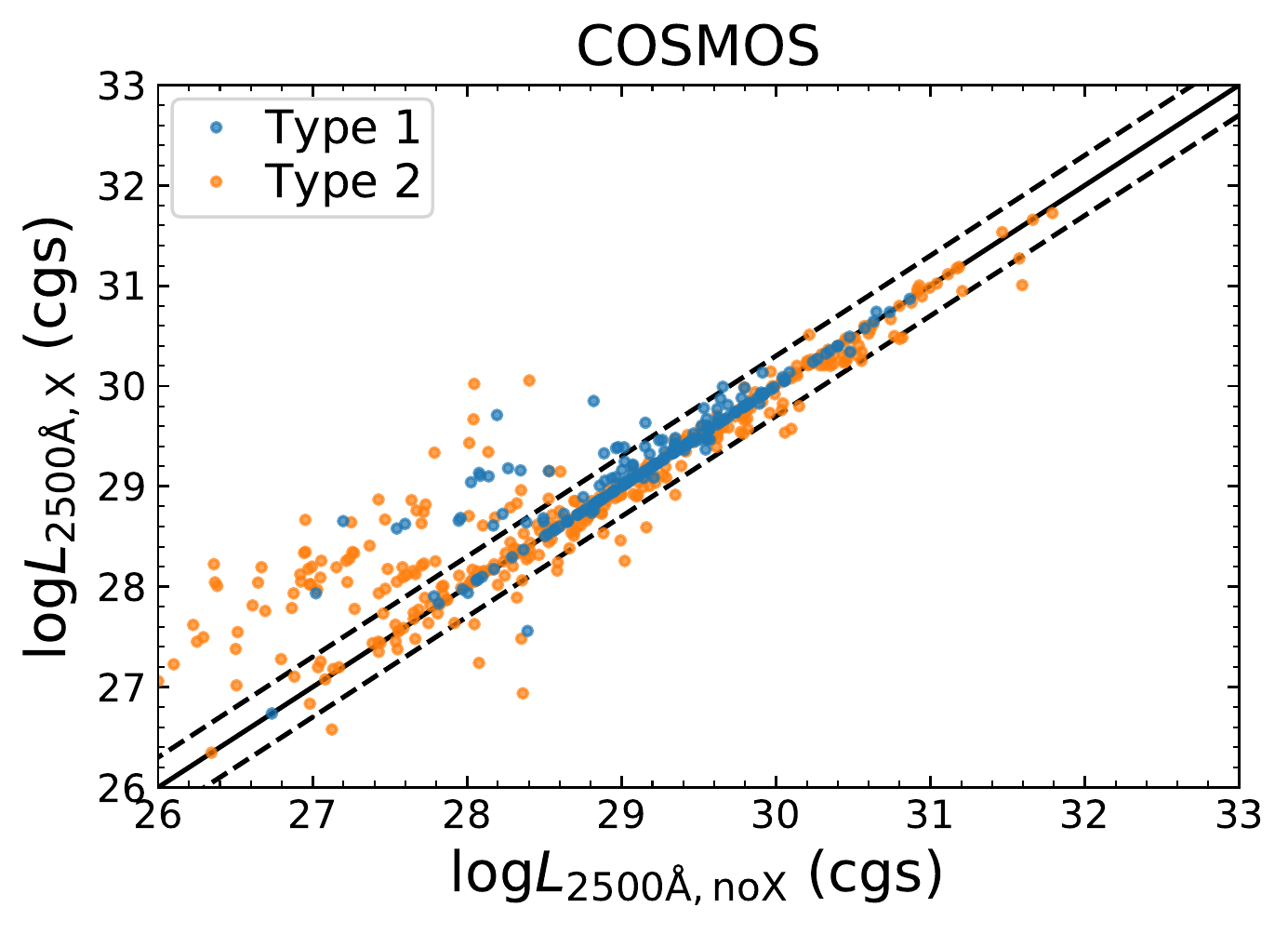}
\includegraphics[width=\columnwidth]{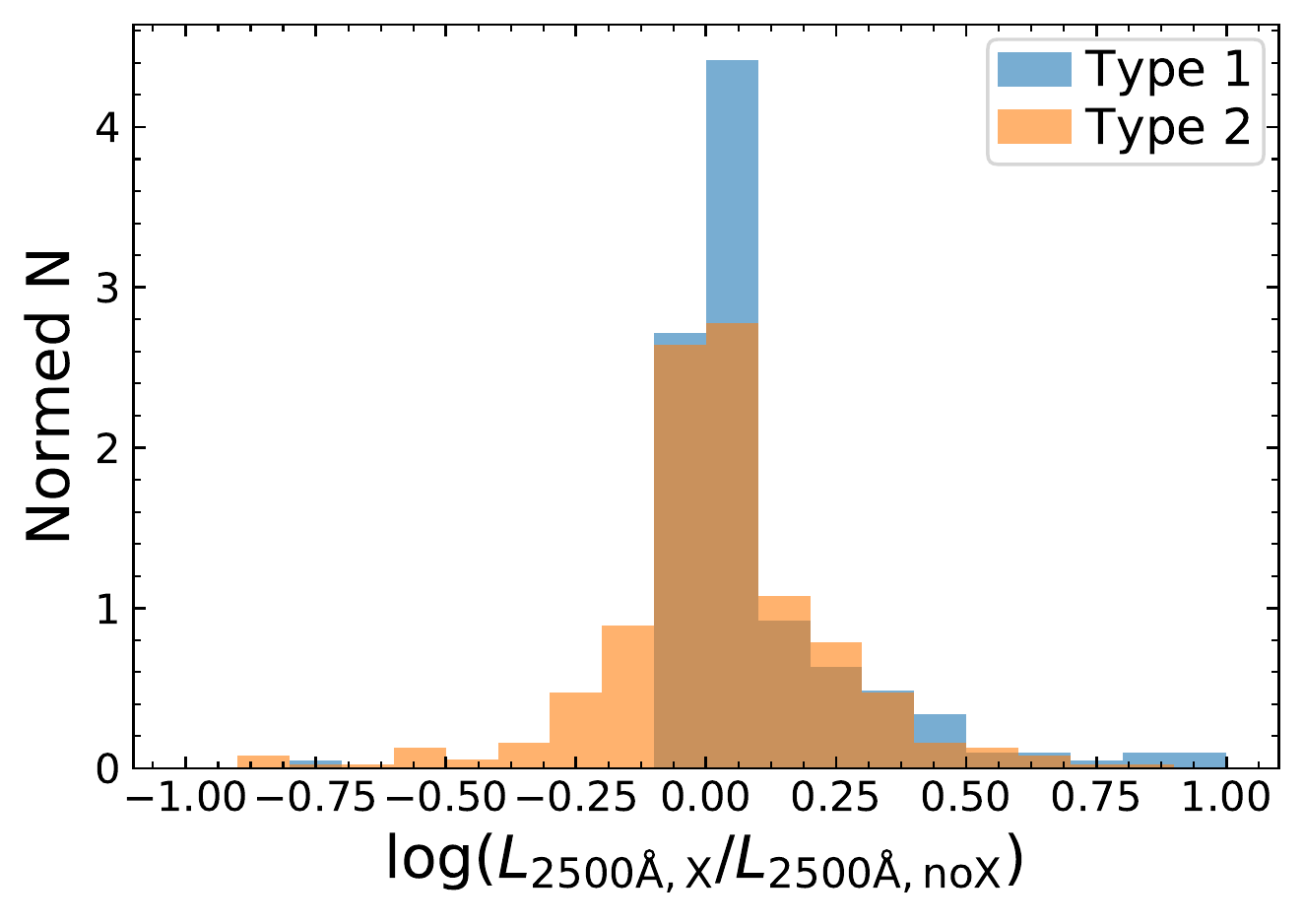}
\caption{Same format as Fig.~\ref{fig:Lbol_vs_Lbol_sdss} but 
for the COSMOS sample.
The blue and orange colors indicate type~1 and type~2 sources,
respectively.
The differences between $\luvx$ and $\luvnox$ are larger 
compared to those in SDSS (Fig.~\ref{fig:Lbol_vs_Lbol_sdss}). 
}
\label{fig:Lbol_vs_Lbol_cosmos}
\end{figure}

\begin{figure}
\includegraphics[width=\columnwidth]{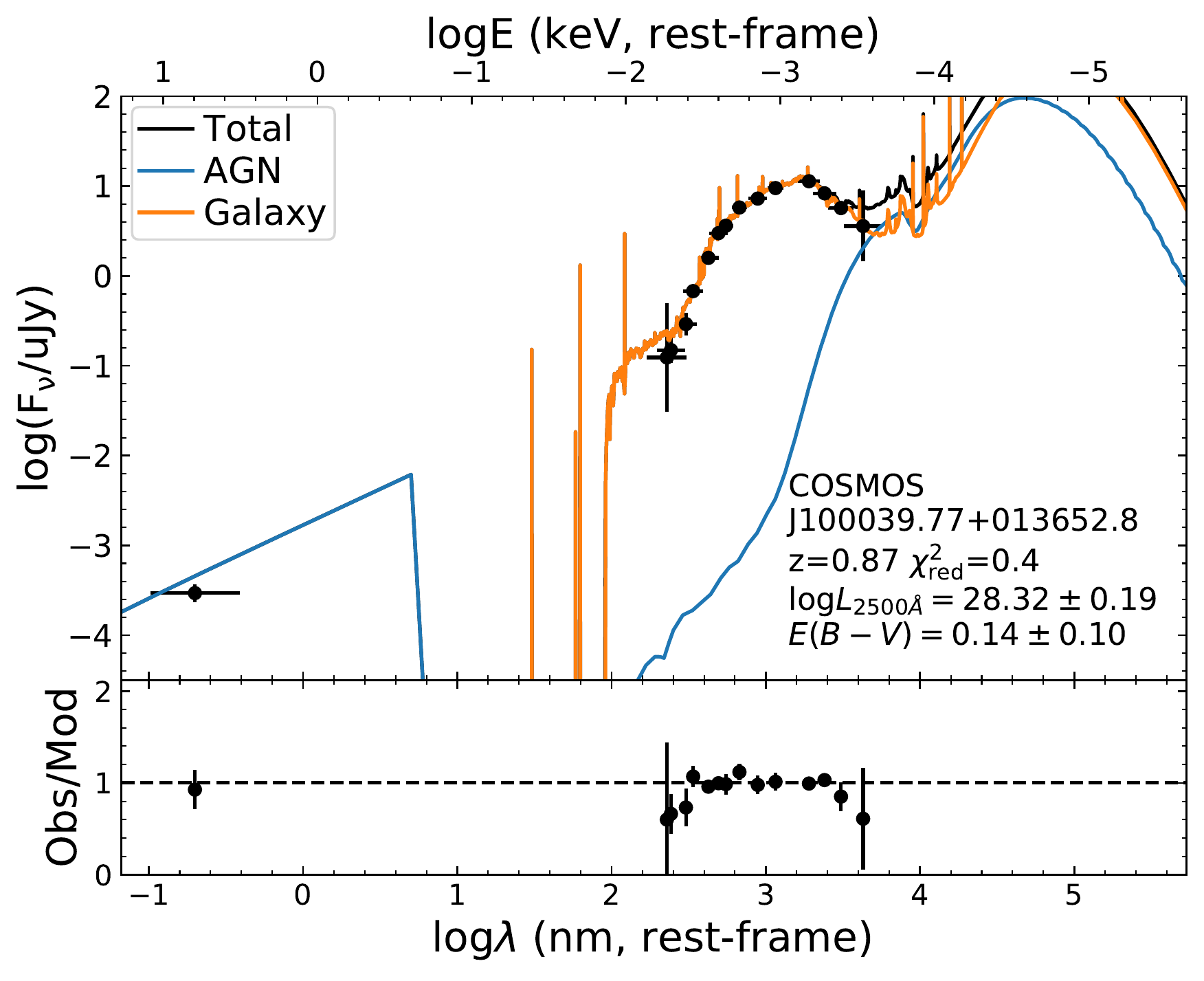}
\includegraphics[width=\columnwidth]{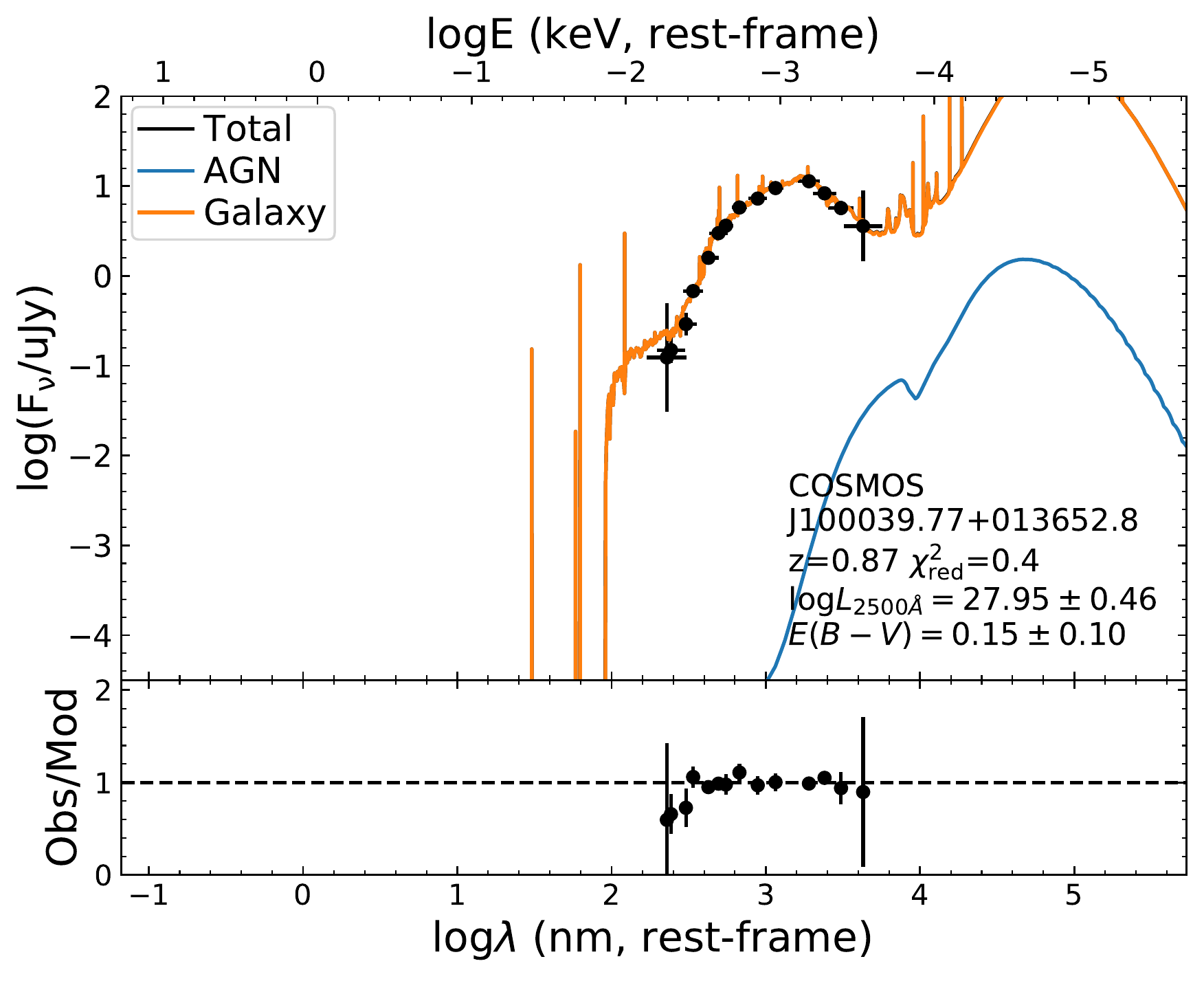}
\caption{An example COSMOS SED fitted with (top) vs.\ without 
(bottom) \xray\ data. 
For this type~2 source, the best-fit intrinsic AGN $\luv$ is 
significantly different in the two cases as labeled.  
The observed UV-to-IR fluxes are dominated by the galaxy
component. 
Therefore, the intrinsic AGN power cannot be effectively 
constrained without the \xray\ data.
}
\label{fig:sed_with_X_cosmos}
\end{figure}

Fig.~\ref{fig:Lbol_vs_Lbol_cosmos} compares $\luvx$ and $\luvnox$ for
the COSMOS sample.
The differences between $\luvx$ and $\luvnox$ are larger 
compared to those in SDSS (Fig.~\ref{fig:Lbol_vs_Lbol_sdss}). 
This is because, for the SDSS sources, the observed optical fluxes are 
dominated by the AGN component, and thus AGN power can be effectively 
constrained even without \xray\ data.
In contrast, for COSMOS, the observed optical-to-IR fluxes are often 
not dominated by AGN, and \xcig\ needs to decompose the fluxes into 
galaxy and AGN components.
This SED decomposition process may be sometimes difficult, given that 
different models could result in similar model fluxes in optical-to-IR
SED.
Therefore, the \xray\ data, which is often dominated by AGN, can be
helpful in constraining the AGN power.
Fig.~\ref{fig:sed_with_X_cosmos} shows an example type~2 AGN SED 
fitted with vs.\ without \xray.
For this source, the observed UV-to-IR fluxes are dominated by the 
galaxy component, and thus AGN power cannot be effectively constrained 
without \xray\ data.

In our fitting (Table~\ref{tab:par_cosmos}), we set viewing angles
at $30^\circ$ and $70^\circ$ for type~1 and type~2 AGNs, respectively.
This model setting can be done for our COSMOS sample, where 
spectroscopic classification is available (\S\ref{sec:sample_cosmos}).
However, for many photometric surveys (e.g. \S\ref{sec:sample_akari}),
the AGN spectra are not available, and thus spectrum-based AGN-type 
classification cannot be performed.
In this case, the \xcig\ user can set both $30^\circ$ and $70^\circ$
for the viewing angle, and allow \xcig\ to freely choose between them.
We test this configuration with our entire COSMOS sample, including 
spectroscopic type~1 and type~2 AGNs.
The other parameters are the same as in Table~\ref{tab:par_cosmos}.
For spectroscopic type~1 (type~2) AGNs, 70\% (28\%) and 30\% (72\%) 
sources have the best-fit viewing angles of $30^\circ$ and $70^\circ$, 
respectively.
This means that, if one uses the best-fit viewing angle to perform 
AGN-type classification (i.e. the SED-based classification), the 
correct rate will be roughly $\approx 70\%$ for both type~1 and 
type~2.

\subsection{AKARI-NEP}\label{sec:akari}

\subsubsection{The sample and the models}\label{sec:sample_akari}
The \hbox{AKARI-NEP} sample is also \xray\ selected ($>3\sigma$, 
\hbox{2--7 keV} band) based on \chandra\ observations of the 
\hbox{AKARI-NEP} field \citep{krumpe15}.
{The \cite{krumpe15} catalog assumes $\Gamma=1.4$. 
Similarly as in \S\ref{sec:sample_sdss} and \S\ref{sec:sample_cosmos}, 
we apply a correction factor of 0.94 (calculated with PIMMS) to 
the \hbox{2--7 keV} fluxes to 
make them consistent with our adopted $\Gamma=1.8$ (AGN).
}
We match the \xray\ with the multiwavelength catalog compiled by 
\cite{buat15} using a $1\arcsec$ matching radius.
This multiwavelength catalog has 19 bands from $u$ to 
\herschel/PACS (100$\mu$m and 160$\mu$m).
Notably, these bands include an excellent set of 9-band MIR data 
from the \akari\ telescope, allowing us to test \xcig\ on the MIR 
wavelengths.
The final sample has 74 sources (Table~\ref{tab:samp_prop}).

For \hbox{AKARI-NEP}, we adopt the same \xcig\ fitting parameters 
as for COSMOS (see Table~\ref{tab:par_cosmos}) except the viewing 
angle.
For the viewing angle, we allow \xcig\ to choose freely between 
30$^\circ$ (type~1) and 70$^\circ$ (type~2), since spectroscopic 
classifications of AGN type are not available for the 
\hbox{AKARI-NEP} sample.

\subsubsection{Fitting results}\label{sec:res_akari}
\begin{figure}
\includegraphics[width=\columnwidth]{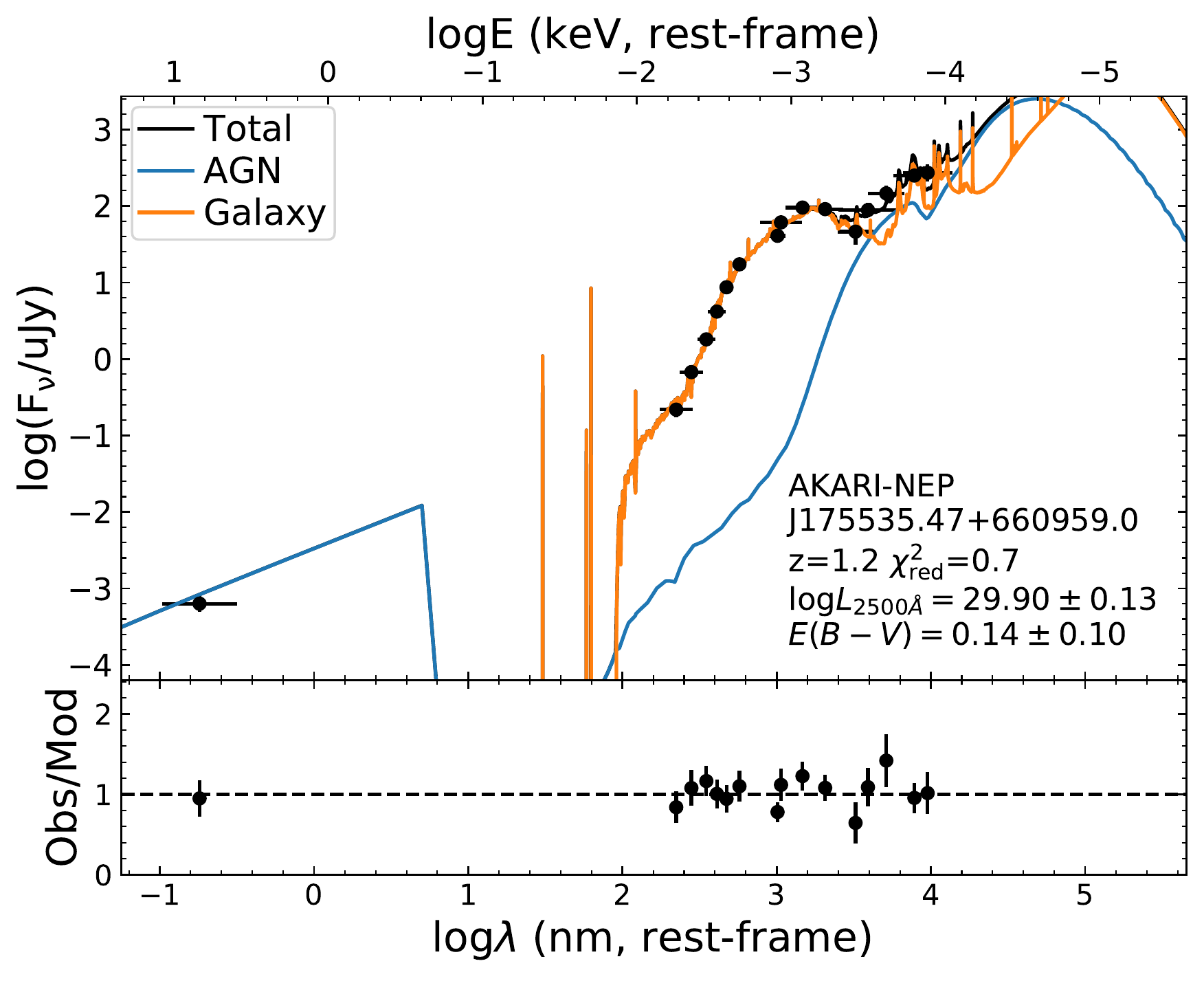}
\includegraphics[width=\columnwidth]{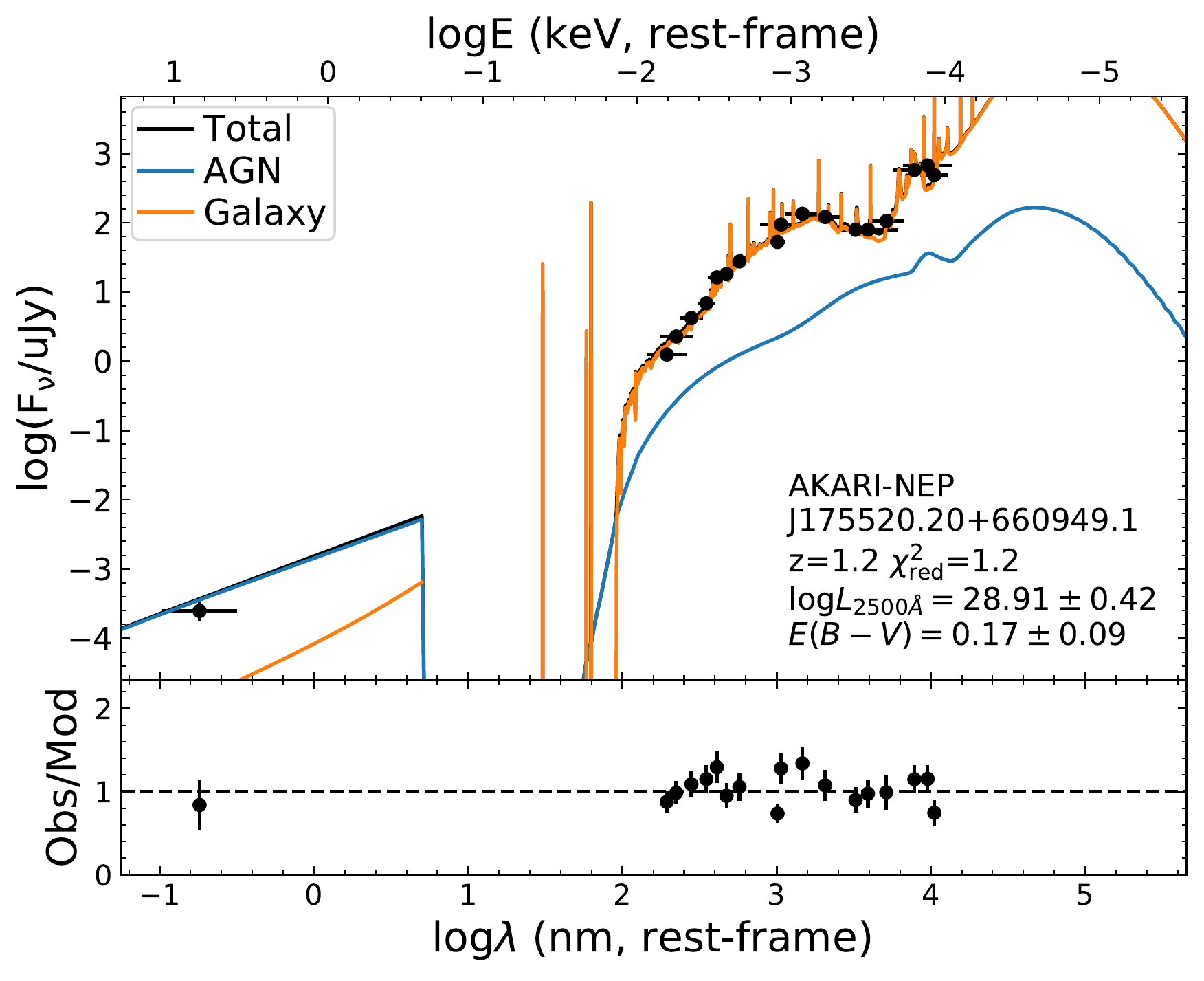}
\caption{Same format as in Fig.~\ref{fig:example_sed_cosmos} 
but for two sources in \hbox{AKARI-NEP}.
Notably, the MIR data are well fitted with our model.
}
\label{fig:example_sed_akari}
\end{figure}

\begin{figure}
\includegraphics[width=\columnwidth]{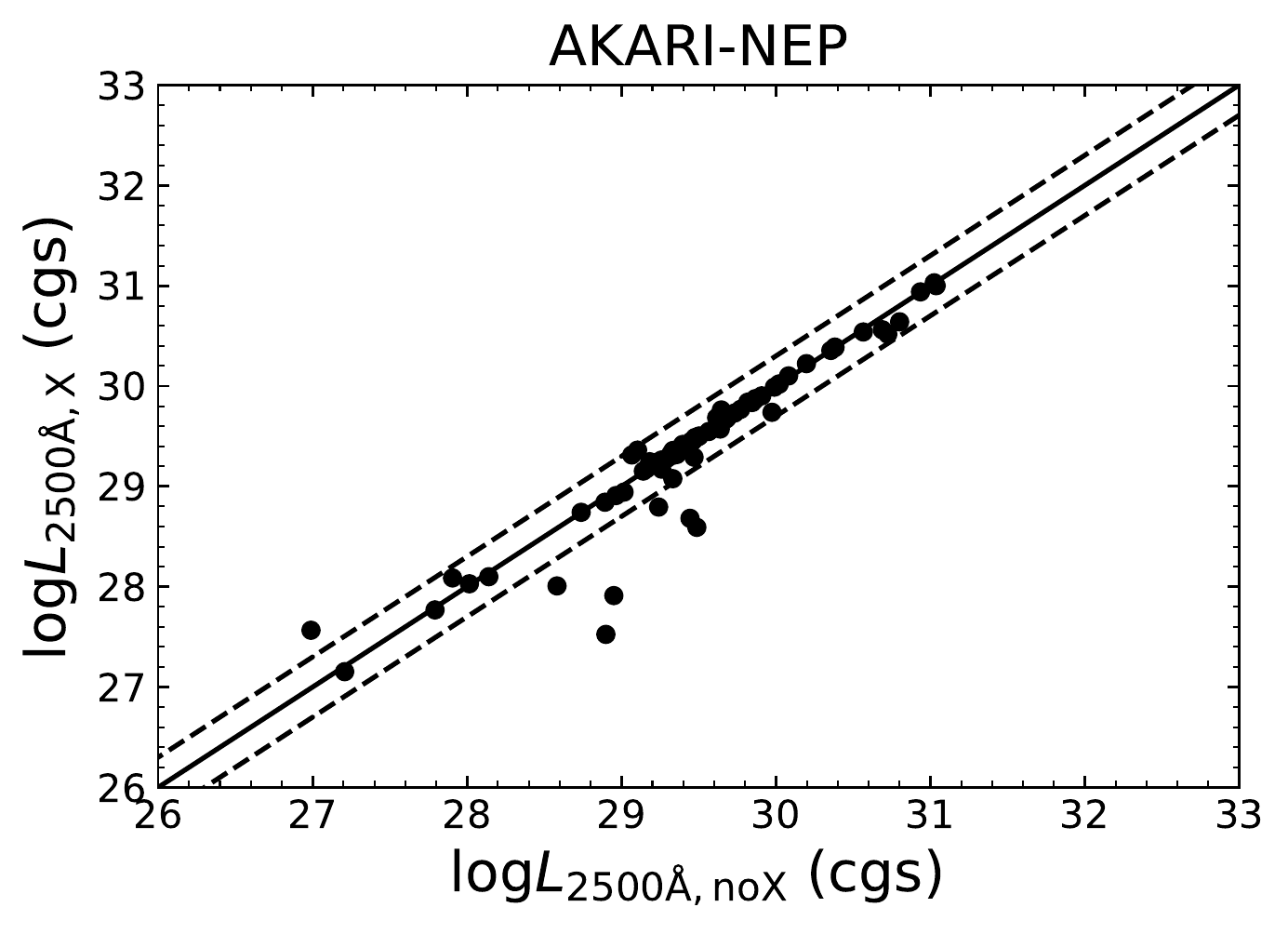}
\includegraphics[width=\columnwidth]{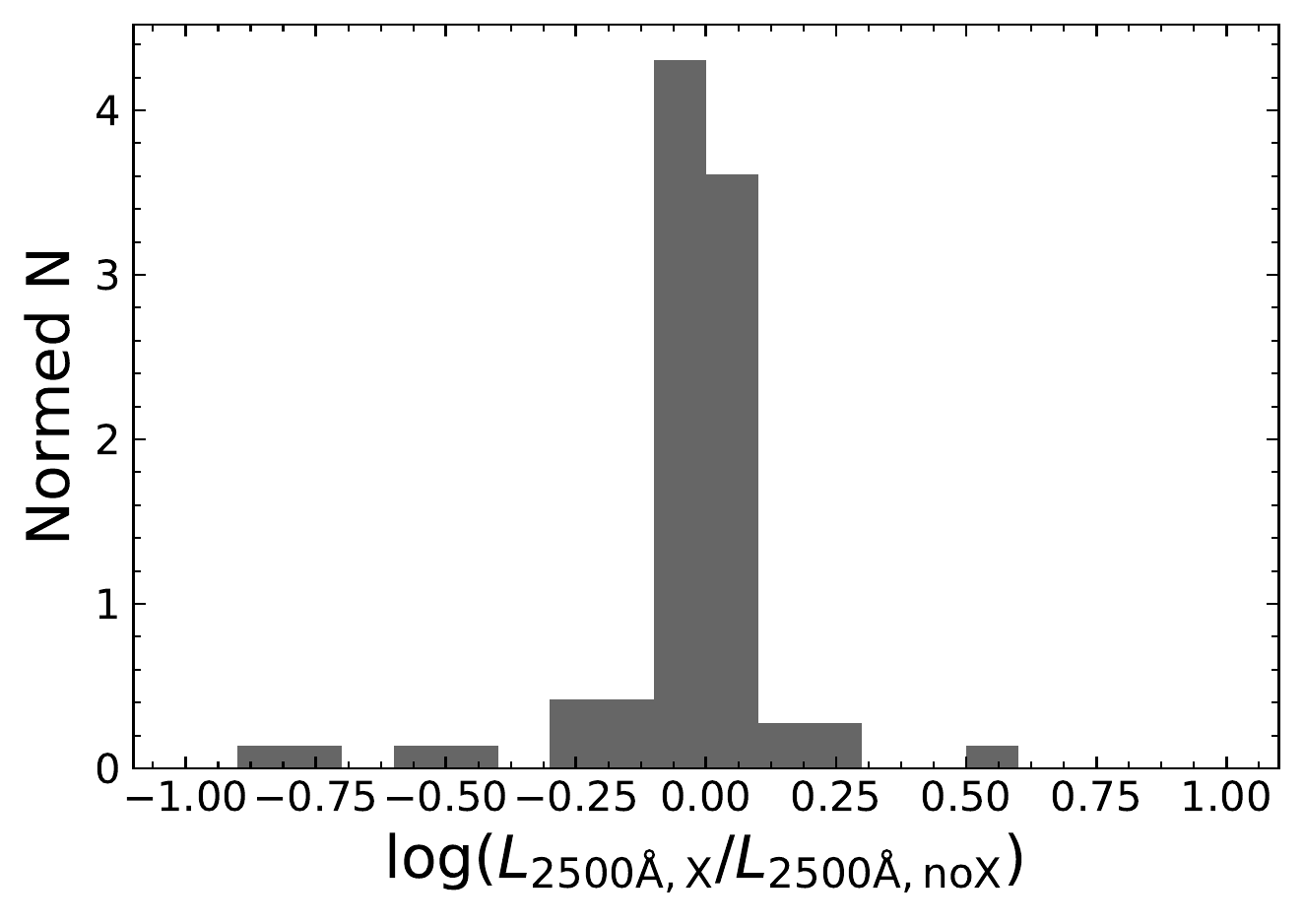}
\caption{Same format as Fig.~\ref{fig:Lbol_vs_Lbol_sdss} but 
for the \hbox{AKARI-NEP} sample.
The differences between $\luvx$ and $\luvnox$ are larger 
compared to those in SDSS (Fig.~\ref{fig:Lbol_vs_Lbol_sdss}). 
}
\label{fig:Lbol_vs_Lbol_akari}
\end{figure}

We run \xcig\ with model parameters in \S\ref{sec:sample_akari}. 
The median $\redchi$ is 1.2 (median dof$=17$).
The resulting $p$-value is 0.27, indicating overall good fitting 
quality.
Fig.~\ref{fig:example_sed_akari} shows two example fitted SEDs
in the \hbox{AKARI-NEP} sample.
Note that the MIR data can be well fitted with our model.
Fig.~\ref{fig:Lbol_vs_Lbol_akari} compares $\luvx$ and $\luvnox$
for \hbox{AKARI-NEP}.
The differences between $\luvx$ and $\luvnox$ are larger 
compared to those in SDSS (Fig.~\ref{fig:Lbol_vs_Lbol_sdss}).
The reason is similar as discussed in \S\ref{sec:res_cosmos},
i.e. SED decomposition is needed for \hbox{AKARI-NEP} and such
decomposition might be ambiguous without \xray.

\subsubsection{Parameter constrainability and degeneracy}\label{sec:deg}
{The \hbox{AKARI-NEP} sample is suitable for investigating the 
constrainability and degeneracy of AGN model parameters, thanks to 
its excellent coverage at MIR wavelengths (\S\ref{sec:sample_akari}) 
where AGN-dust emission peaks (Fig.~\ref{fig:sed_ebv}).
In this section, we present discussion on model-parameter
constrainability and degeneracy using the \hbox{AKARI-NEP} sample.
Since the population in this sample is AGN-galaxy mixed systems 
in general (e.g. Fig.~\ref{fig:example_sed_akari}), the results 
below might not be applicable to some particular sources, for which 
the observed MIR emission is dominated by AGNs (e.g. hot dust-obscured 
galaxies, hot DOGs; \hbox{\citealt{dey08}}; \hbox{\citealt{vito18b}}).
For these AGN-dominated sources, the AGN parameters might be 
easier to constrain, as host-galaxy contributions to the 
MIR emission are negligible.
}

\begin{figure*}
\includegraphics[width=\columnwidth]{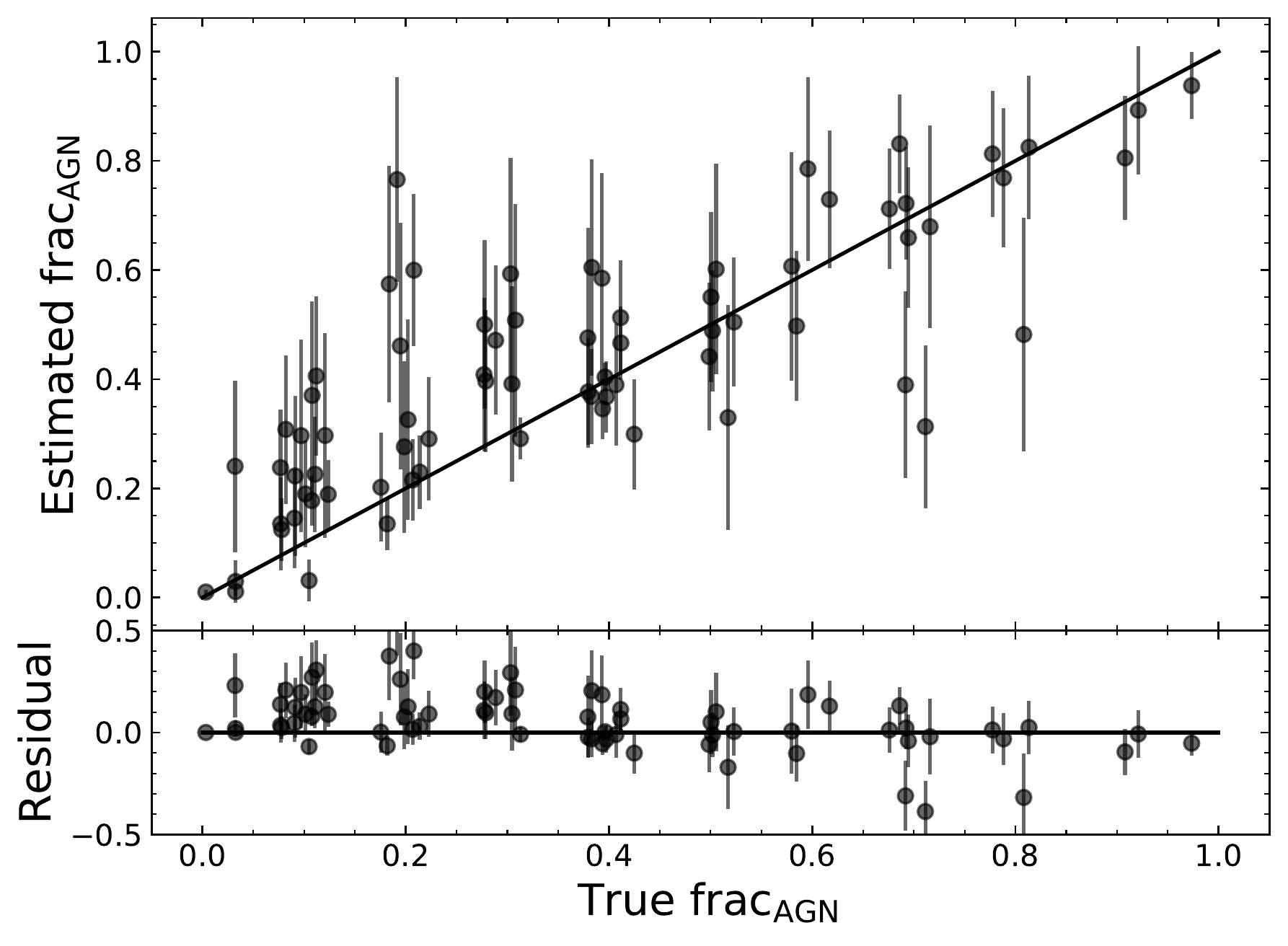}
\includegraphics[width=\columnwidth]{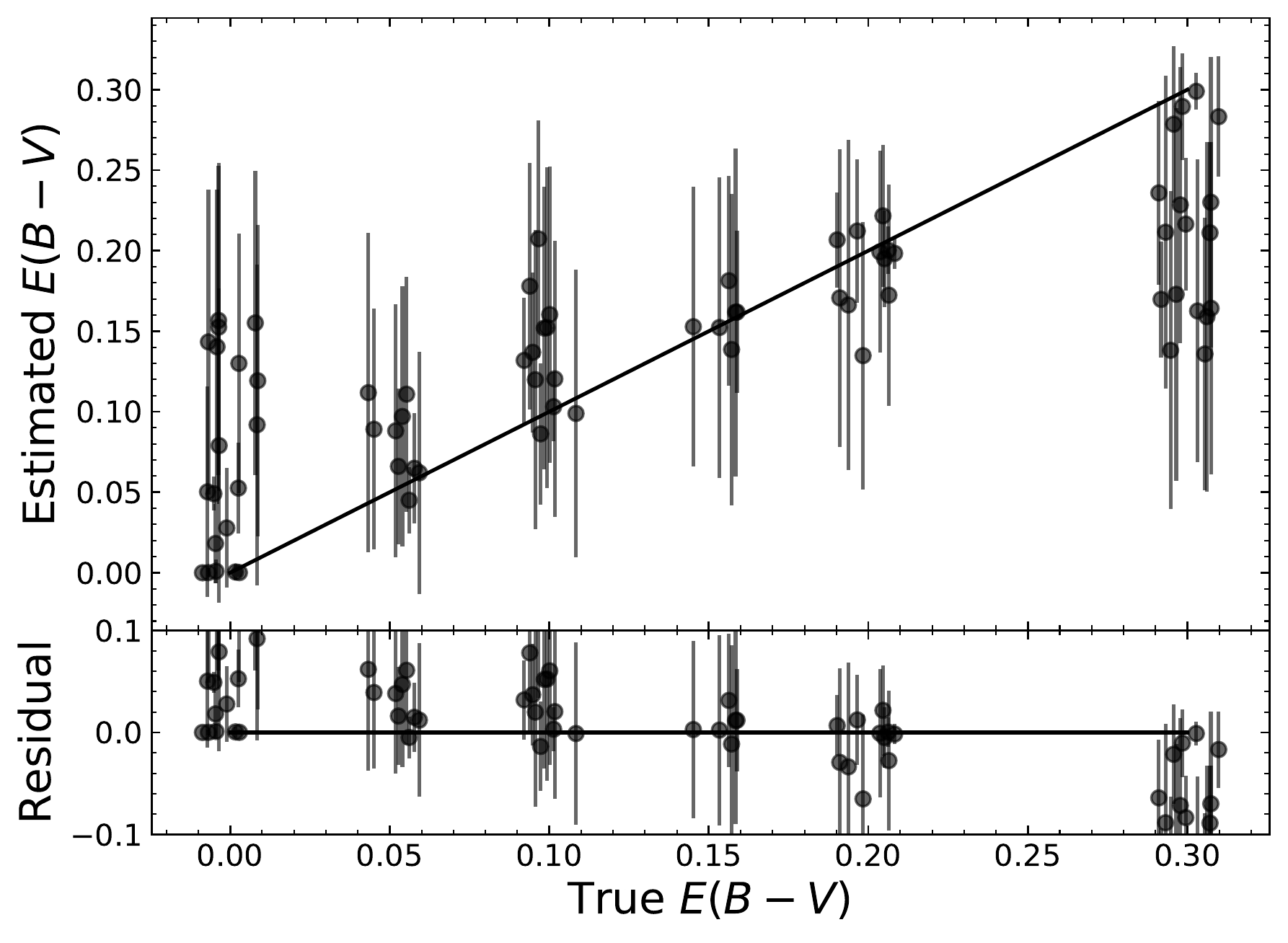}
\caption{Same format as in Fig.~\ref{fig:mock_sdss} but for
$\fracA$ (left) and $E(B-V)$ (right) of the \hbox{AKARI-NEP} AGNs.
}
\label{fig:mock_frac_akari}
\end{figure*}

{There are three parameters that have multiple values in 
our fitting (Table~\ref{tab:par_cosmos}), i.e. viewing angle, 
$\fracA$, and polar-dust $E(B-V)$.
The viewing angle determines AGN types (see \S\ref{sec:polar_mod}),
and we show that the spectroscopic AGN types can be recovered 
with $\approx 70\%$ accuracy in \S\ref{sec:res_cosmos}. 
For $\fracA$ and $E(B-V)$, we run the mock analysis as a 
sanity check of their constrainability (\S\ref{sec:res_sdss}), 
and the results are displayed in Fig.~\ref{fig:mock_frac_akari}.
For $\fracA$, the estimated and true values are generally 
correlated (median errors $= 0.13$), although some sources 
have relatively large uncertainties. 
Therefore, the relative IR emission strength between AGN and 
galaxy (as measured by $\fracA$) can be effectively constrained.
In contrast, unlike the case of SDSS (\S\ref{sec:res_sdss}),
the estimated $E(B-V)$ is relatively flat as a function 
of true $E(B-V)$, indicating that $E(B-V)$ cannot be well 
constrained in general.
This result is understandable.
For SDSS, the SED is dominated by type~1 AGNs, and the $E(B-V)$ 
is directly related to the observed UV/optical SED shape.
However, for \hbox{AKARI-NEP}, the SEDs are generally produced 
by both AGN and galaxy components, and the $E(B-V)$ cannot 
be determined directly from the UV/optical SED shape (or other 
SED features). 
}

\begin{figure}
\includegraphics[width=\columnwidth]{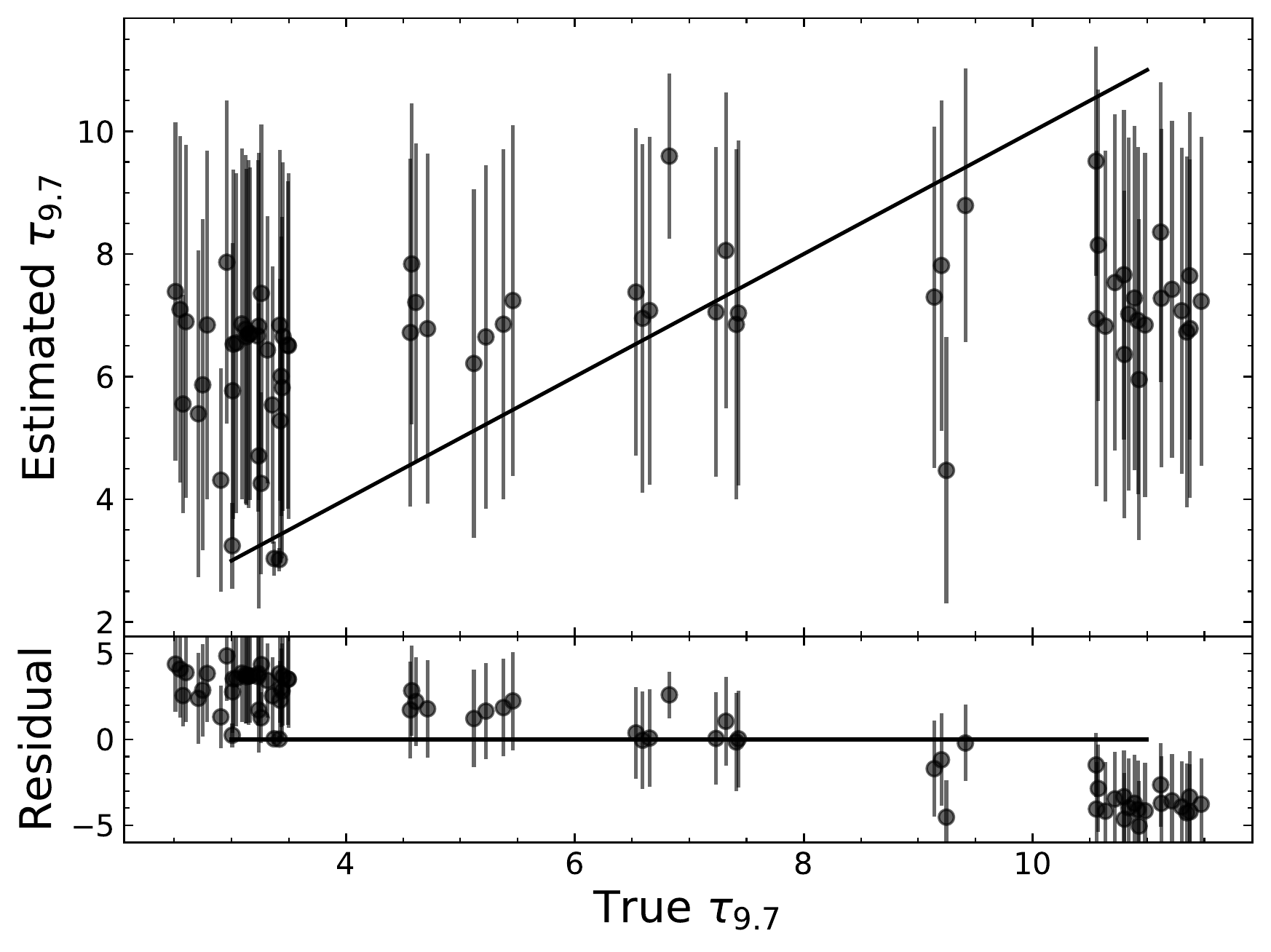}
\caption{Same format as in Fig.~\ref{fig:mock_sdss} but for
torus $\tau_{9.7}$ of the \hbox{AKARI-NEP} AGNs.
For $\tau_{9.7}$, the estimated value is relatively 
flat as a function of the true value, indicating that they cannot
be well constrained by the observed data.
}
\label{fig:mock_tau_akari}
\end{figure}

{In our fitting (Table~\ref{tab:par_cosmos}), most 
of the torus and polar-dust parameters such as $\tau_{9.7}$ and 
polar-dust temperature are fixed at single values. 
This is because these parameters are related to the MIR SED 
shape. 
Considering that model degeneracy is likely strong in the MIR, 
broad-band photometry data like \hbox{AKARI-NEP} might not 
be able to effectively constrain these parameters.
Now, we test whether torus $\tau_{9.7}$ can be well constrained.
The \xcig\ configuration is the same as in 
Table~\ref{tab:par_cosmos} except that $\tau_{9.7}$ is allowed to
vary among 3, 5, 7, 9, and 11 (all allowed values).
The mock-analysis results are presented in 
Fig.~\ref{fig:mock_tau_akari}.
The estimated value is generally flat as a function of the 
true value, indicating that $\tau_{9.7}$ cannot be well constrained.
We have also tested other fixed AGN parameters in 
Table~\ref{tab:par_cosmos} such as polar-dust 
temperature and torus opening angle, and found they 
cannot be effectively constrained either.
}

\begin{figure*}
\includegraphics[width=\columnwidth]{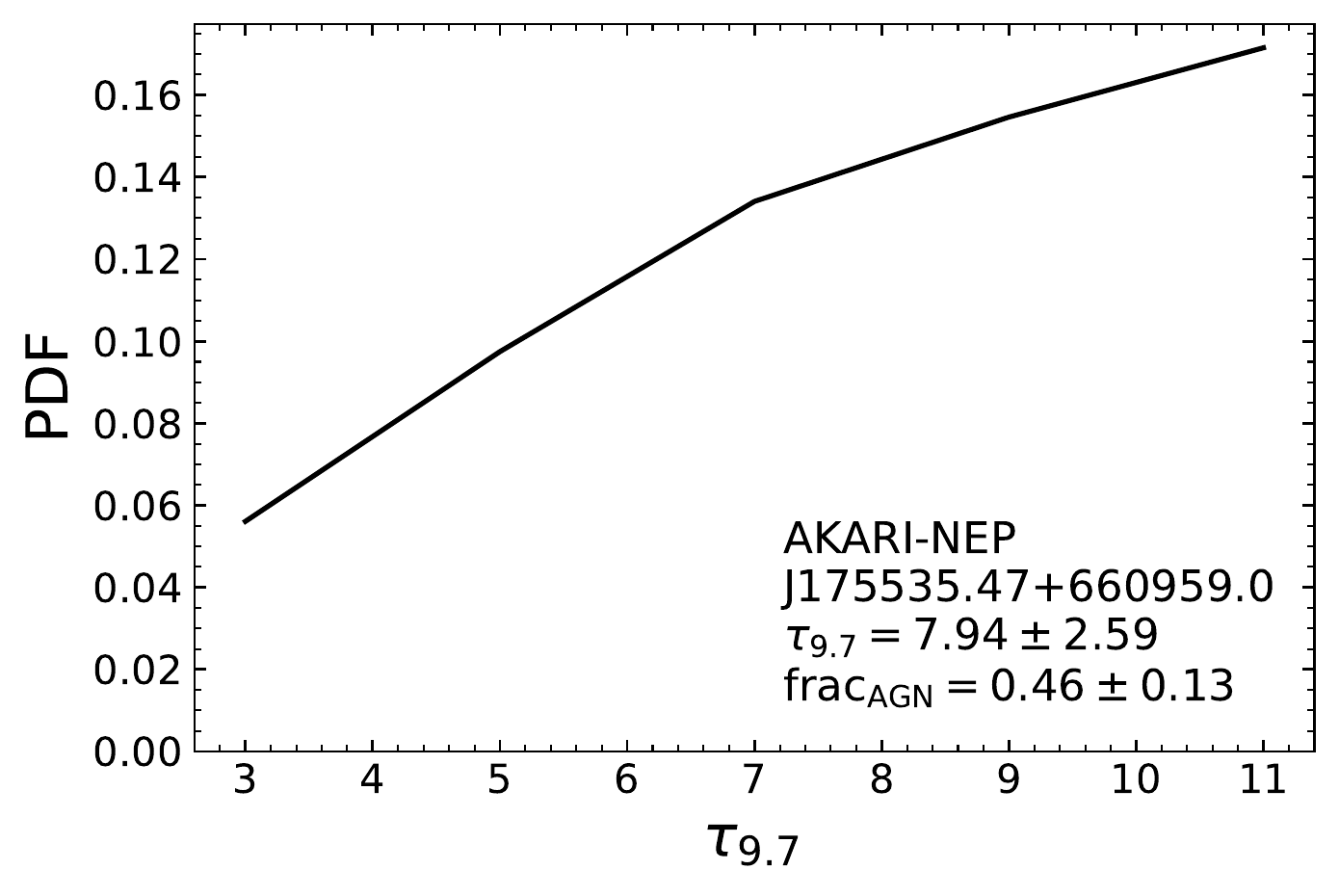}
\includegraphics[width=\columnwidth]{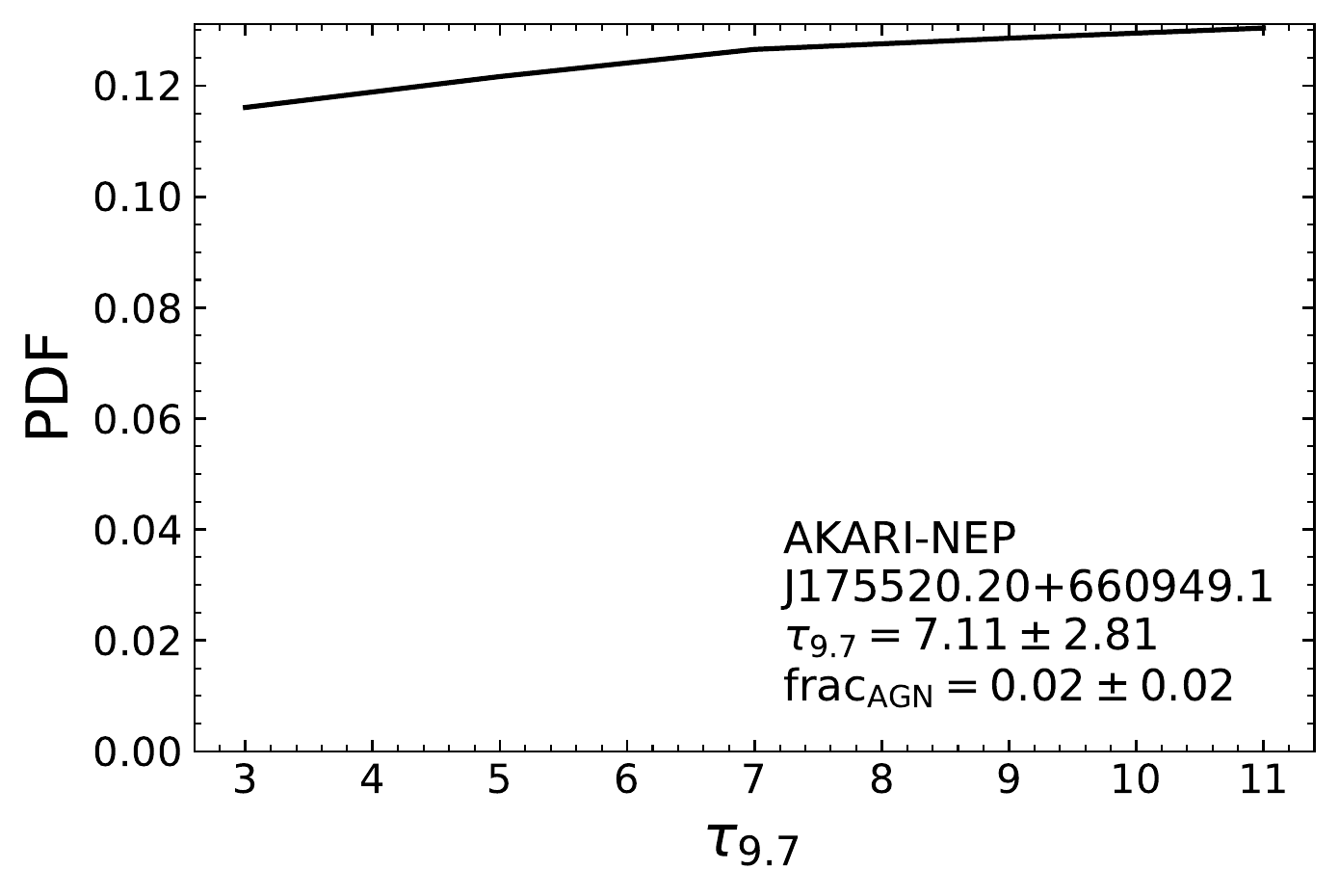}
\includegraphics[width=\columnwidth]{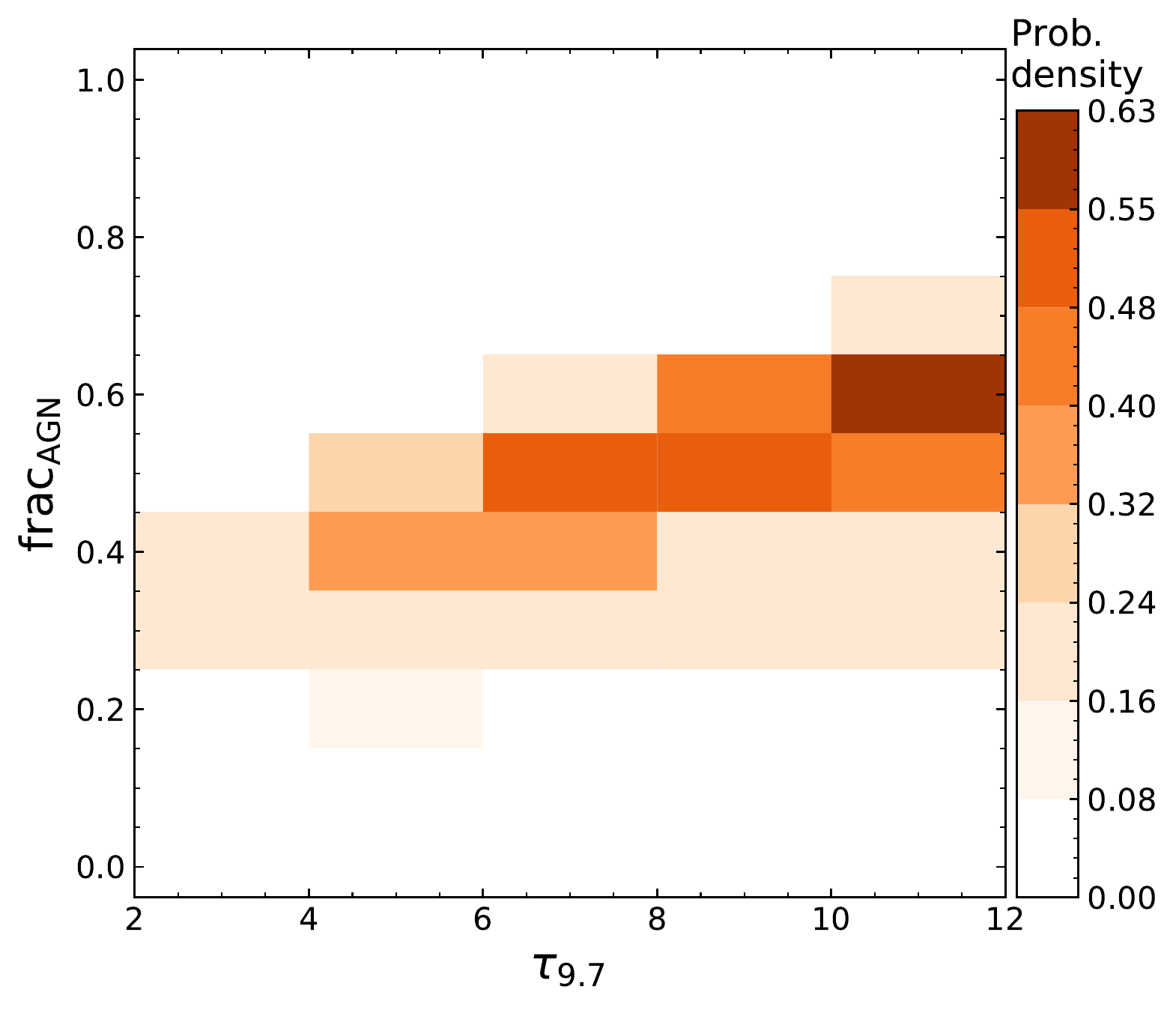}
\includegraphics[width=\columnwidth]{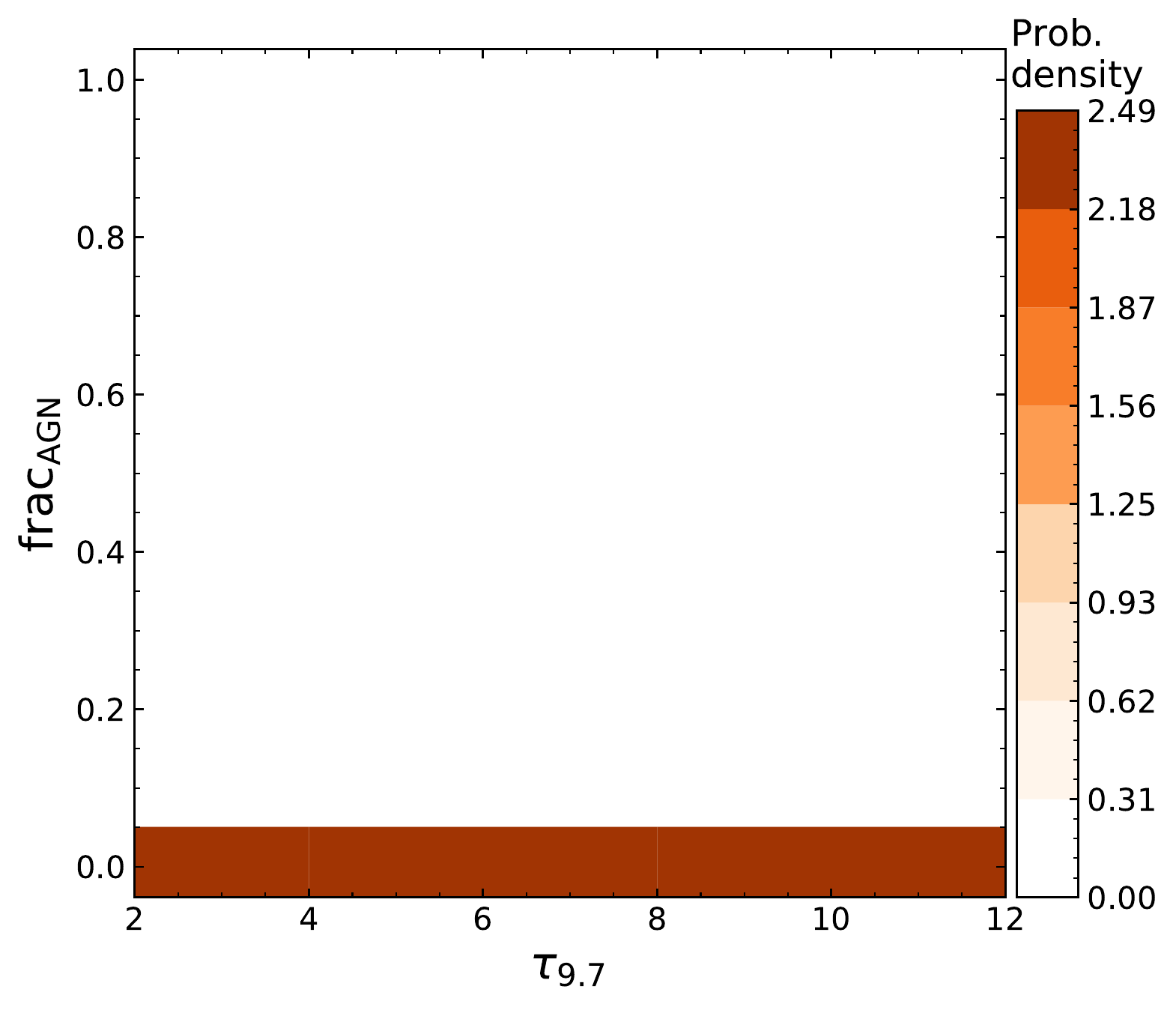} \\
\caption{{Top: the $\tau_{9.7}$ PDF for the example source in 
Fig.~\ref{fig:example_sed_akari}.
Bottom: the 2D probability density map of $\tau_{9.7}$ vs.\ 
$\fracA$.
The density map is normalized such that the 2D integral 
is unity.
}}
\label{fig:example_pdf_akari}
\end{figure*}

{
Now, we analyze the reasons why $\tau_{9.7}$ has generally 
large uncertainties, using the two sources in 
Fig.~\ref{fig:example_sed_akari} as illustrative examples.
These reasons also generally explain the large uncertainties 
of other unconstrained parameters.
In Fig.~\ref{fig:example_pdf_akari}, we show the $\tau_{9.7}$ PDFs 
and 2D probability density maps ($\tau_{9.7}$ vs.\ $\fracA$).
For J175535.47+660959.0, the $\fracA$ can be constrained to
$0.46\pm0.13$.  
From the density map, at high $\fracA$ ($\approx 0.6$), 
the probability peaks at $\tau_{9.7}=11$. 
However, at lower $\fracA$, 
the peak shifts to lower $\tau_{9.7}$.
Therefore, the $\tau_{9.7}$ parameter is degenerate with $\fracA$,
i.e. the probability distribution of $\tau_{9.7}$ depends on the
value of $\fracA$.
This degeneracy makes the marginalized $\tau_{9.7}$ PDF relatively 
flat, leading to the large uncertainty of this parameter.
From the density map of J175520.20+660949.1, $\fracA$ is well
constrained at a low level, but the $\tau_{9.7}$ PDF is flat.
This is because when $\fracA$ is low, the observed MIR SED is 
dominated by the galaxy component (Fig.~\ref{fig:example_sed_akari} 
bottom). 
In this case, the models with different $\tau_{9.7}$ are 
degenerate, in the sense that they have similar MIR SED shapes
due to dominant galaxy contributions.
The results above indicate that model degeneracy is responsible for 
the large uncertainties in the $\tau_{9.7}$ estimation.
}

\section{Fitting with X-ray upper limits}\label{sec:uplim}
In the previous section, we apply \xcig\ to sources with \xray\ detections.
The \xray\ sources are often only a small fraction ($\lesssim 10\%$)
of the entire sample in optical/IR surveys.
However, many extragalactic studies need to constrain AGN emission for the 
\xray\ undetected majority galaxy population 
(e.g. \hbox{\citealt{buat15}}; \hbox{\citealt{vito16}}; 
\hbox{\citealt{bowman19}}).
This task can be achieved by \xcig, based on \xray\ flux upper limits.
The detailed fitting algorithm is presented in \S4.3 of \cite{boquien19},
and we do not repeat it here.
Below, we test this usage with 100 randomly selected \hbox{AKARI-NEP} 
galaxies \citep{buat15}, all of which are \xray\ undetected. 
Here, we choose \hbox{AKARI-NEP} rather than SDSS or COSMOS, since 
\hbox{AKARI-NEP} has the best multiwavelength coverage among the three 
surveys (\S\ref{sec:test}).
We focus on a relatively small sample (100), because the upper-limit 
analysis in \xcig\ is time-consuming due to its complicated mathematical 
form of $\chi^2$ (see \S4.3 of \citealt{boquien19}).

Precise \xray\ flux upper limits depend on source positions and vary from 
source to source. 
The derivation of the precise values requires intensive simulations 
\citep[e.g.][]{xue16, luo17}, which are beyond the scope of this work.
Instead, we adopt a single conservative value, 
5.3$\times 10^{-15}$~erg~s$^{-1}$~cm$^{-2}$ (4.4$\times 10^{-7}$~mJy, 
\hbox{2--7}~keV), for all the sources.
For $\approx 80\%$ of the survey area, the actual \chandra\ flux limits should 
be lower than this value (see Fig.~10 of \hbox{\citealt{krumpe15}}).
We run \xcig\ based on this upper limit. 
The parameter settings are the same as in \S\ref{sec:sample_akari} except 
that we allow $\fracA=0$, since it is possible that the AGN component does 
not exist for these upper-limit sources.
For comparison, we also re-run \xcig\ without the \xray\ upper limit.

\begin{figure}
\includegraphics[width=\columnwidth]{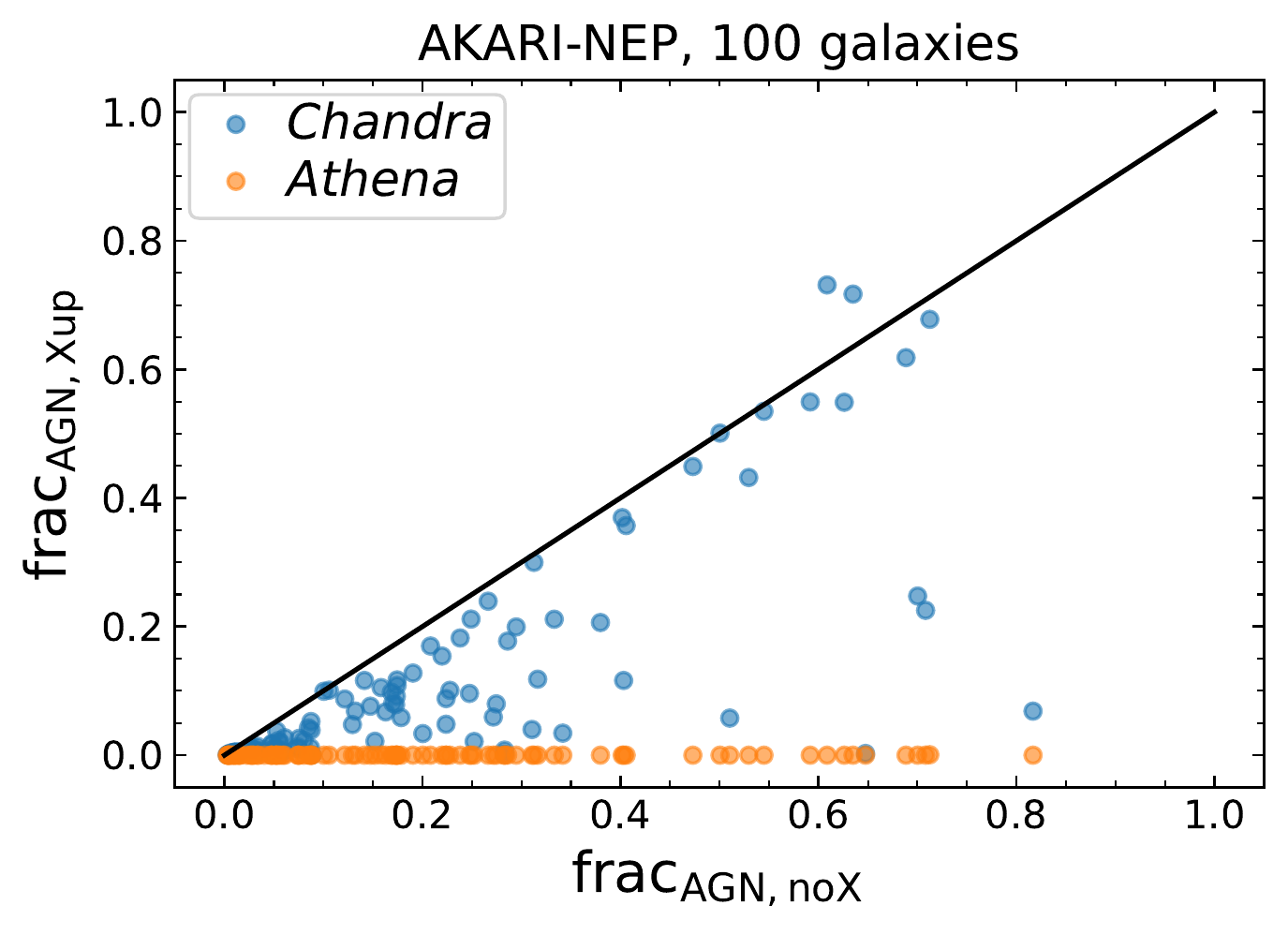}
\caption{Top: Comparison of $\fracA$ between the fitting without 
vs.\ with \xray\ upper limits for 100 random galaxies in the 
\hbox{AKARI-NEP} field.
The blue and orange points represent $\fracAx$ obtained from 
\chandra\ and \athena\ upper limits, respectively.
The solid black lines indicate the 1:1 relation.
As expected, $\fracAx$ is generally lower than $\fracAnox$, since
the \xray\ upper limit can constrain AGN power. 
The current \chandra\ upper limit can effectively constrain AGN 
power for some sources.
The \athena\ upper limit strongly suppresses $\fracAx$ to $<1\%$
for all sources.
}
\label{fig:frac_vs_frac}
\end{figure}

Fig.~\ref{fig:frac_vs_frac} compares the {Bayesian-like} estimate of $\fracA$ 
of these two runs ($\fracAnox$ vs.\ $\fracAx$). 
As expected, $\fracAx$ is generally lower than $\fracAnox$, since
the \xray\ upper limit can constrain AGN power. 
Notably, for $\approx 10\%$ of the sources, the $\fracAx$ is much lower 
($\Delta\fracAx > 0.2$) than $\fracAnox$.
This result indicates that, even when excellent MIR data are present
(\S\ref{sec:sample_akari}), the AGN-galaxy decomposition might still 
be inaccurate/ambiguous without \xray\ data.   
On the other hand, there are still $\approx 30\%$ of sources that have 
non-negligible $\fracA$ ($>0.1$) when the \xray\ upper limit is used
in the fitting.
It is possible that a non-negligible IR flux is contributed by the AGN.
However, another possibility is that the current \xray\ upper limit 
is too high to effectively constrain AGN emission.
In the future, \athena\ may clarify this problem with its great 
sensitivity.

The total \chandra\ exposure time on the \hbox{AKARI-NEP} field is 
$\approx 300$~ks. 
Given this amount of exposure time, \athena\ can reach its 
confusion-limited sensitivity of 
$\sim 1\times 10^{-16}$~erg~s$^{-1}$~cm$^{-2}$ ($8 \times 10^{-9}$~mJy, 
2--7~keV) for the entire \hbox{AKARI-NEP} 
field.\footnote{\chandra\ can, in principal, reach deeper 
sensitivity than \athena\ thanks to its superior angular resolution.
However, reaching \athena-like (or deeper) flux limits will practically 
need large amounts of exposure time of \chandra.
This has only been achieved in two small \chandra\ fields (only 
$\sim 500$~arcmin$^2$ each), i.e. 7~Ms \hbox{CDF-S} and 
2~Ms \hbox{CDF-N} \citep{xue16, luo17}.
}
Assuming this \xray\ flux limit, we re-run \xcig. 
The resulting $\fracAx$ is below 0.01 for all sources  
(see Fig.~\ref{fig:frac_vs_frac}), indicating that the AGN 
SED contribution will be negligible if a source is undetected by 
\athena.
Therefore, \xcig, with future \athena\ observations, will have great 
power in unambiguously determining the presence of AGN.
This feature will be extremely helpful for future extragalactic 
studies.

\section{Summary and future prospects}\label{sec:sum}
We have developed and tested \xcig, a new version of the galaxy SED 
fitting code, CIGALE. 
Our development and test results are summarized below.
\begin{enumerate}

\item We have developed a new \xray\ module (\S\ref{sec:xray_mod}).
The module is mainly designed to connect the intrinsic \xray\ 
emission with other wavelengths, and \xray\ obscuration and 
transmission should be corrected before providing the \xray\ data 
to \xcig.
The \xray\ module includes the \xray\ emission from both galaxy 
and AGN. 
The galaxy component includes the emission from HMXB, LMXB, and 
hot gas.
The AGN's \xray\ SED is connected to its UV-to-IR SED using the 
well-known $\ox-\luv$ relation.

\item We have implemented a modern torus model, SKIRTOR, to fit 
AGN UV-to-IR SEDs (\S\ref{sec:skirtor}).
SKIRTOR adopts a clumpy two-phase torus, which is 
responsible for obscuring the UV/optical emission from the AGN 
disk.
SKIRTOR is developed from a 3D radiative-transfer method, and 
thus obeys the energy-conservation law.
However, SKIRTOR assumes that the AGN disk emission is absolutely 
unextincted when viewed from the polar direction. 
Therefore, SKIRTOR cannot {model} the SEDs of slightly extincted
type~1 AGNs. 
To overcome this disadvantage, we introduce extinction from 
polar dust (\S\ref{sec:polar_dust}).
The extinction amplitude, $E(B-V)$, is a free model parameter
set by the \xcig\ user. 

\item We have tested \xcig\ on the AGNs with \xray\ detections
in SDSS, COSMOS, and \hbox{AKARI-NEP} \S\ref{sec:test}.
The three samples have distinctive characteristics in terms of 
AGN properties and available data.
The fitting quality is good in general, with typical 
$\chi^2 \sim 1$ for all the samples. 
This result indicates that \xcig\ is capable in {modelling} 
observed AGN SEDs under different circumstances.
We also compare the fittings results with vs.\ without \xray\ data.
We find that the resulting AGN power is sometimes different 
in the two cases, when both AGN and galaxy components are 
present.
Therefore, the AGN-galaxy SED decomposition may be unphysical
without the constraints from \xray\ data.
{We discuss constrainability and degeneracy of model 
parameters in the fitting of \hbox{AKARI-NEP}, for which excellent
mid-IR photometric coverage is available.
}

\item We also test \xcig\ on a random sample of galaxies 
with only \chandra\ \xray\ upper limits in the \hbox{AKARI-NEP} 
field, where excellent MIR data are available (\S\ref{sec:uplim}).
We compare the fitting results with and without the \xray\ 
upper limits.
After using the \xray\ upper limits, $\fracA$ sometimes 
becomes lower, indicating that the current \chandra\ upper 
limit can effectively constrain AGN emission, as least for 
some systems.
We also evaluate the potential of the future \athena\ mission 
by replacing the \chandra\ upper limit with the expected \athena\
value for a similar exposure time ($\approx 300$~ks).
The resulting $\fracA$ is constrained to a negligible level 
($<1\%$) for all the sources, indicating that \athena\ can 
robustly constrain AGN emission in general with a moderate 
amount of exposure time ($\lesssim 300$~ks). 

\end{enumerate}

We publicly release \xcig\ on the official website of 
CIGALE.\footnote{https://cigale.lam.fr/}
As for the previous versions of CIGALE, \xcig\ is 
open-source, allowing the user to modify the source code
freely.
In the future, we will further develop \xcig\ and enable
it to address special AGNs such as radio-loud and BAL 
objects (\S\ref{sec:xray_mod}).
Besides the three surveys tested in this work 
(SDSS, COSMOS, and \hbox{AKARI-NEP}),
the user can apply \xcig\ to the existing multiwavelength 
surveys such as \hbox{CDF-S} \citep{luo17}, \hbox{CDF-N} 
\citep{xue16}, and \hbox{XMM-SERVS} \citep{chen18}.
In the future, \xcig\ can be used to explore deep/wide 
surveys of, e.g. \erosita\ \citep[e.g.][]{merloni12} and 
\athena\ \citep[e.g.][]{nandra13}.

\section*{Acknowledgments}
We thank the referee for helpful feedback that improved
this work.
GY acknowledges support from the French Space Agency 
CNES.
VB received funding from Excellence Initiative of 
Aix-Marseille University - AMIDEX, a French 
``Investissements d'Avenir'' programme.
MS acknowledges support by the Ministry of Education, Science,
and Technological Development of the Republic of Serbia through
the projects Astrophysical Spectroscopy of Extragalactic Objects
(176001) and Gravitation and the Large Scale Structure of the
Universe (176003).
WNB acknowledges support from 
NASA ADP grant 80NSSC18K0878 and the V.M. Willaman Endowment.
This project uses Astropy (a Python package; see 
\citealt{astropy}).




\bibliographystyle{mnras}
\bibliography{all.bib} 



\appendix

\section{New inputs and outputs in \xcig}
\label{sec:in_and_out}
The input parameters for the new \xray\ and SKIRTOR modules can 
be found in Table~\ref{tab:par_cosmos}.
After fitting, \xcig\ can output the best-fit model SEDs
of different components, and the SED components for the new \xray\ 
and SKIRTOR modules are summarized in Table~\ref{tab:out_sed}.
{Besides the best-fit SEDs, \xcig\ can also yield the 
maximum-likelihood and {Bayesian-like} values of source physical properties. 
These physical properties include not only the model parameters 
(Table~\ref{tab:par_cosmos}), but also some additional quantities 
as listed in Table~\ref{tab:out_phy}.}
New quantities can be added in the future as requested by the 
user of \xcig.
We remind that the quantities, ``agn.intrin\_Lnu\_2500A'' and 
``xray.alpha\_ox'' refers to the values as measured at a viewing 
angle of $30^\circ$ (see \S\ref{sec:ox}). 
The quantity ``agn.accretion\_power'' refers to the intrinsic 
(unextincted) AGN disk luminosity averaged over all directions
(weighted by $\sin\theta$; see \S\ref{sec:ox}).
This quantity, paired with an assumed radiative efficiency, 
can be used to estimate BH accretion rate (e.g. 
\hbox{\citealt{yang17, yang19}}).

\begin{table*}
\centering
\caption{Output SED components for the new \xray\ and SKIRTOR modules}
\label{tab:out_sed}
\begin{tabular}{lll} \hline\hline
Module & Component & Explanation \\
\hline
    \multirow{2}{*}{\shortstack[l]{X-ray (\S\ref{sec:xray_sed}) \\ $10^{-3}$--5~nm}} 
    & xray.agn & The AGN corona\\
    & xray.galaxy & The total {SED} of HMXB, LMXB, and hot gas\\
\hline
    \multirow{2}{*}{\shortstack[l]{SKIRTOR (\S\ref{sec:skirtor}) \\ 8--10$^6$~nm}} 
    & agn.SKIRTOR2016\_disk & The AGN disk\\ 
    & agn.SKIRTOR2016\_dust & The dust reemission\\
\hline
\end{tabular}
\begin{flushleft}
{\sc Note.} --- {In \xcig\ output, all the SED components are in the format 
                of $L_{\lambda}$ in units of W~nm$^{-1}$.}
\end{flushleft}
\end{table*}

\begin{table*}
\centering
\caption{Additional output physical parameters for the 
        new \xray\ and SKIRTOR modules}
\label{tab:out_phy}
\begin{tabular}{llll} \hline\hline
Module & Parameters & Explanation & Units \\
\hline
    \multirow{7}{*}{\shortstack[l]{X-ray}} 
    & xray.agn\_Lnu\_2keV & The AGN $L_\nu$ at 2~keV & W Hz$^{-1}$ \\
    & xray.agn\_Lx\_2to10keV & The AGN 2--10~keV luminosity & W \\
    & xray.agn\_Lx\_total & The AGN total (0.25--1200~keV) \xray\ luminosity & W \\
    & xray.alpha\_ox & The AGN $\ox$ & $-$ \\
    & xray.lmxb\_Lx\_2to10keV & The 2--10~keV LMXB luminosity & W \\
    & xray.hmxb\_Lx\_2to10keV & The 2--10~keV HMXB luminosity & W \\
    & xray.hotgas\_Lx\_0p5to2keV & The 0.5--2~keV hot-gas luminosity & W \\
\hline
    \multirow{5}{*}{\shortstack[l]{SKIRTOR}} 
    & agn.disk\_luminosity & The observed AGN disk luminosity (might be extincted) & W \\ 
    & agn.dust\_luminosity & The observed AGN dust reemitted luminosity & W \\
    & agn.luminosity & The sum of agn.disk\_luminosity and agn.dust\_luminosity & W \\
    & agn.intrin\_Lnu\_2500A & The intrinsic AGN $L_\nu$ at 2500~\AA\ at
                               viewing angle $=30^\circ$ & W Hz$^{-1}$ \\
    & agn.accretion\_power & The intrinsic AGN disk luminosity averaged over all directions & W \\
\hline
\end{tabular}
\end{table*}

%


\bsp	
\label{lastpage}
\end{document}